\documentclass[lettersize,journal]{IEEEtran}
\usepackage{amsmath,amsfonts}
\usepackage{algorithmic}
\usepackage{algorithm}
\usepackage{array}
\usepackage[caption=false,font=normalsize,labelfont=sf,textfont=sf]{subfig}
\usepackage{textcomp}
\usepackage{stfloats}
\usepackage{url}
\usepackage{verbatim}
\usepackage{graphicx}
\usepackage{cite}
\usepackage{algorithmic}
\usepackage{graphicx}
\usepackage{textcomp}
\usepackage{xcolor}
\usepackage{multirow}
\usepackage{pifont}
\usepackage{tabularx}
\usepackage{booktabs}
\usepackage{threeparttable}
\usepackage{enumitem}
\begin{document}

\title{FPGN: Redefining Ultra-Fast Programmable Gate-based Neural Acceleration with Differentiable LUTs}

\author{Jiawei Liang$^1$, Haotong Qin$^2$, Linfeng Du$^1$, Xingyu Liu$^1$, Shangkun Li$^1$, Hui Yu$^1$, \\Michele Magno$^2$, Xinyu Chen$^3$, Jiang Xu$^3$, Wei Zhang$^1$\\
$^1$HKUST, $^2$ETH Zurich, $^3$HKUST (GZ)
}



\maketitle

\begin{abstract}
Achieving nanosecond-scale inference latency for deep neural networks (DNNs) has become a primary architectural concern for latency-critical applications. While Field-Programmable Gate Arrays (FPGAs) offer a promising substrate for low-latency inference, conventional FPGA accelerators remain arithmetic-centric, using LUTs primarily as building blocks for numerical operators and peripheral logic. In contrast, recent LUT-native neural networks treat LUTs as learnable neurons, revealing promising theoretical potential to exploit their intrinsic logic expressivity. However, existing methods are largely confined to algorithmic optimizations, failing to translate this theoretical potential into high-performance FPGA accelerators. Specifically, their differentiable formulations do not faithfully match FPGA LUT primitives, their physically-unaware topologies compromise routability and timing closure, and their lack of automated optimization flow hinders systematic design space exploration (DSE) and efficient hardware implementation.

In this paper, we propose FPGN, an end-to-end physically-aware framework that closes the gap between LUT-native learning and latency-optimized FPGA implementation. FPGN addresses these challenges through (i) a hardware-aligned differentiable formulation for training FPGA-native LUT neurons, (ii) a structured LUT-native topology with a streaming hardware architecture to improve routing locality and timing closure, and (iii) a latency-driven compiler that leverages high-fidelity analytical Quality of Results models to automate DSE and hardware generation. Experiments show that FPGN achieves up to 205$\times$ latency reduction compared to representative FPGA-based BNN accelerators and up to 30$\times$ higher LUT efficiency than prior differentiable LUT-native networks, while maintaining competitive inference accuracy. 
\end{abstract}

\begin{IEEEkeywords}
Differentiable, FPGA, LUT-Native Networks, Hardware Co-Design.
\end{IEEEkeywords}

\section{Introduction}\label{sec:intro}
Deep Neural Networks (DNNs) have driven major advances in artificial intelligence \cite{vaswani2017attention,he2016deep}. However, deploying them in domains prioritizing nanosecond-scale response requirements over other constraints remains highly challenging, such as high-energy physics triggers \cite{guo2025klinq,le2025computing}, high-frequency trading \cite{jia2022domain,yoo2023lighttrader}, and line-rate packet or flow classification in high-speed network data planes \cite{swamy2022taurus,jafri2024leo}. While GPUs dominate high-throughput training, their rigid memory hierarchy and batch-oriented execution make them architecturally unsuitable for these latency-critical applications \cite{boutros2025field}. 

Specialized accelerators on Field-Programmable Gate Arrays (FPGAs) \cite{zhang2015optimizing,ghasemzadeh2018rebnet} and Application-Specific Integrated Circuits (ASICs) \cite{ramachandran2025microscopiq,ding20252} have been developed to reduce inference latency. While ASICs offer peak performance, their inflexibility and high Non-Recurring Engineering (NRE) costs limit their adaptability to rapidly evolving algorithms \cite{nurvitadhi2019compete}. Consequently, FPGAs have emerged as a compelling substrate for hardware-algorithm co-design, offering agile reconfigurability alongside deterministic ultra-low latency. At the core of this flexibility is the $k$-input Look-Up Table ($k$-LUT), a small memory that can be programmed to realize any Boolean function of $k$ input bits. 

\begin{figure}
    \centering
    \includegraphics[width=0.7\linewidth]{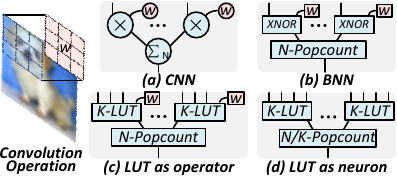}
    \caption{Evolution of neural paradigms on FPGA. (a) MAC in CNN. (b) BNNs replace MAC with XNOR and N-input popcount. (c) The LUT-as-operator paradigm replaces XNOR with learnable $k$-LUTs while maintaining arithmetic-centric structures. (d) LUT-as-neuron paradigm unifies operators and weights into learnable LUTs.}
    \label{fig_LUTNet_example}
\end{figure}

Despite inherent programmability, traditional FPGA accelerators follow an arithmetic-centric paradigm that treats LUTs as building blocks to implement predefined arithmetic and glue logic. For instance, Convolutional Neural Network (CNN) accelerators on FPGA \cite{zhang2015optimizing,lei2019acceleration} utilize LUTs and Digital Signal Processing (DSP) slices to implement Multiply-Accumulate~(MAC) as shown in Fig.~\ref{fig_LUTNet_example}(a). While parallelizable, the heavy memory access overhead and computational complexity of these models remain barriers to ultra-low latency implementation. Binary Neural Networks (BNNs) \cite{rastegari2016xnor,ghasemzadeh2018rebnet} mitigate this issue by binarizing weights to one-bit values and replacing time-consuming MACs with efficient bit-wise XNOR and population count~(popcount) operations (Fig.~\ref{fig_LUTNet_example}(b)). However, they ultimately confine LUTs to executing numerical functions, leaving the Boolean expressivity of LUTs largely underutilized.

More promising methods make LUTs themselves learnable, following two distinct paradigms. The first treats LUTs as learnable operators, as exemplified by LUTNet \cite{wang2020lutnet}. It replaces XNOR gates with LUTs to increase local flexibility (Fig.~\ref{fig_LUTNet_example}(c)). However, this operation-level replacement still inherits the legacy BNN datapath and weights, which imposes structural constraints forcing the learned LUTs to remain functionally proximal to XNOR \cite{wang2020lutnet}. The second more radical paradigm \cite{andronic2023polylut, andronic2024neuralut, andronic2025neuralut} instead unifies operators and weights into learnable LUTs as illustrated in Fig.~\ref{fig_LUTNet_example}(d), thereby eliminating the weight-induced memory access overhead. While effective, these approaches typically rely on predefined function to fill LUT configurations, which explore a negligible fraction of the LUT’s massive $2^{2^k}$ Boolean functional space.

Recent algorithmic advances \cite{petersen2022deep, petersen2024convolutional, bacellar2024differentiable} have begun to construct logic-native neural networks directly on differentiable logic. Prior works \cite{petersen2022deep,petersen2024convolutional} pioneered the use of two-input logic gates as neurons, successfully pushing the latency of the image classification task on the CIFAR-10~\cite{CIFAR10} dataset down to the nanosecond level while maintaining accuracy competitive with representative BNNs. DWN~\cite{bacellar2024differentiable} expands this idea to 6-LUTs that is most widely used in modern FPGAs to balance performance and area efficiency~\cite{boutros2021fpga}. Orthogonal to these classic connection-centric CNN/MLP formulations, alternative activation-centric paradigms such as KANEL\'E~\cite{hoang2026kanele} have emerged to tabulate univariate splines for scientific computing.

However, existing LUT-native methods remain largely confined to algorithmic optimizations. They either rely on low-input logic elements (e.g., 2-input gates)~\cite{petersen2022deep,petersen2024convolutional} that underutilize FPGA LUT primitives or lack the training scalability to support more complex networks beyond small MLPs~\cite{bacellar2024differentiable}. More importantly, by permitting unstructured connectivity to enhance network performance, they neglect the routing and timing constraints of physical FPGA fabric, making it difficult to realize high-frequency, ultra-low-latency implementations. In addition, the absence of LUT-centric compiler support further hinders automated deployment of such designs. \textbf{This reveals our key insight: transforming the FPGA fabric itself into a learnable LUT-native neural network unlocks the potential for ultra-low-latency neural acceleration, and realizing the potential in practical nanosecond-scale FPGA hardware requires full-stack co-optimization of differentiable LUT training, physically-aware topology, streaming architecture, and compilation.}

Guided by these insights, we propose FPGN, a physically-aware differentiable LUT-native framework for Ultra-\underline{\textbf{F}}ast \underline{\textbf{P}}rogrammable \underline{\textbf{G}}ate-based \underline{\textbf{N}}eural accelerators. It is designed to address three core challenges: (i) the mismatch between differentiable training and discrete FPGA LUT primitives, (ii) the topology-architecture co-design problem imposed by FPGA physical constraints, and (iii) the absence of an automated optimization flow for LUT-native networks. To this end, FPGN is organized as a bottom-up co-design framework with three tightly coupled components:

\begin{itemize}
    \item \textbf{Hardware-Aligned training methodology}: FPGN first introduces a training methodology for LUT-native networks leveraging a differentiable relaxation for LUT neurons coupled with a progressive binarization process. This formulation enables stable convergence and efficient optimization within the massive $2^{2^k}$ Boolean functional space, thereby bridging continuous domain of gradient-based optimization and discrete nature of LUT-native hardware.
    \item \textbf{Physically-Aware Topology and Streaming Architecture}: Building upon the trainable LUT neurons, FPGN proposes a physically-aware LUT-native topology with structured connectivity and a fully streaming FPGA architecture to realize it. By co-designing connectivity, dataflow, and adaptive pipelining, the proposed approach ensures superior physical routability and timing closure to sustain high-frequency nanosecond-scale neural inference.
    \item \textbf{Latency-Driven Automated Compiler}: To instantiate the proposed architecture into optimized streaming accelerators, FPGN integrates high-fidelity analytical Quality of Results (QoR) models into a latency-driven compiler that automatically explores the design space and generates optimized FPGA implementations.
    \item \textbf{An Open-Source Toolchain:} To automate the end-to-end flow from model training to FPGA deployment, we develop an automated toolchain, which is publicly available at \url{https://github.com/MasLiang/FPGN}.
\end{itemize}

\section{Background and Motivation}\label{sec:background}
\subsection{Arithmetic-Centric Paradigm}
Traditional FPGA-based CNN accelerators are inherently arithmetic-centric, necessitating frequent weight fetching to perform computationally intensive multi-bit MAC operations $z=\sum_{i=0}^{n-1}w_i \cdot a_i$ (Fig.~\ref{fig_LUTNet_example}(a)). This incurs heavy memory access overhead and long computational latencies, representing fundamental barriers to ultra-low-latency hardware implementation.

To reduce memory cost and inference latency, BNNs binarize both weights and activations to bit values in $\{0, 1\}$, thereby transforming the computationally expensive MAC operation into a sequence of efficient bit-wise XNOR and popcount operations:
\begin{equation}
z_b = \operatorname{popcount}\big(\operatorname{XNOR}(\mathbf{w}^b,\mathbf{a}^b)\big)
\end{equation}
where $\mathbf{w}^b\in\{0, 1\}$ and $\mathbf{a}^b\in\{0, 1\}$ are vectors of binary values. 

A BNN neuron is illustrated in Fig.~\ref{fig_LUTNet_example}(b). When mapped onto FPGAs, such as in the AMD FINN framework~\cite{umuroglu2017finn}, this arithmetic-centric paradigm treats the LUTs merely as primitives for synthesizing numerical logic operations comprising XNOR gates and adders. Even with physical optimizations incorporating LUT packing \cite{ahmed2009packing} (e.g., merging multiple XNOR and corresponding popcount into a single 6-LUT), a $k$-LUT is confined to only a tiny fraction of its total functional space. This functional underutilization forces the network to rely on increased depth or width to compensate for the limited expressiveness of individual neurons, inevitably increasing total inference latency.

\subsection{LUT-as-Operator Paradigm}
To better exploit FPGA logic, LUTNet \cite{wang2020lutnet} introduces learnable LUTs into pretrained BNNs through an interpolation-based differentiable formulation, as illustrated in Fig.~\ref{fig_LUTNet_example}(c). However, this paradigm is limited by a structural mismatch between the single-bit output of a physical LUT and BNNs' arithmetic datapath. In a standard BNN, multiple XNOR results are aggregated by a popcount operation, which produces results in the non-negative integer domain. In contrast, a $k$-LUT is strictly constrained to a single-bit output. As a result, when a $k$-LUT ($k>2$) is used to replace multiple XNOR operations, it is forced to compress a multi-bit aggregate into a single-bit signal before the popcount stage. This mismatch restricts the expressivity of learned LUTs and largely constrains the optimization to converge toward the original BNN patterns. 
Ultimately, this structural mismatch prevents the paradigm from effectively exploring the vast Boolean function space inherent to the underlying LUT primitive.

\subsection{LUT-as-Neuron Paradigm}
A more radical paradigm treats LUTs as standalone neurons as illustrated in Fig.~\ref{fig_LUTNet_example}(d). Unlike the operator-level approach, these methods use LUT configuration bits to encode both neural parameters and logic operations and utilize specific network topology, enabling a more direct mapping to the FPGA's fundamental logic fabric. LTN \cite{petersen2022deep} and DLN \cite{petersen2024convolutional} rely solely on two-input logic gates without auxiliary operators such as popcount. They enumerate all possible Boolean functions of a gate to represent a neuron as a differentiable weighted combination of these functions. Training updates these weights, after which the Boolean function with the largest weight is selected. Achieving accuracy comparable to FINN on CIFAR-10 image classification, these works lower the latency to nanosecond-level. Another method DWN \cite{bacellar2024differentiable} designs a discrete gradient function by extending finite difference~\cite{strikwerda2004finite}, marking the first time this paradigm is scaled to 6-LUT neurons that align with modern FPGAs. These advancements inspire us to transform the FPGA fabric itself into a learnable LUT-native neural network to unlock the full potential for nanosecond-scale neural acceleration.

\subsection{Challenges}\label{sec_challenge}
While recent advances reveal the potential of using FPGAs as learnable neural substrates, porting this algorithmic success to high-performance hardware still faces three fundamental challenges.

\noindent\textbf{\textit{Challenge 1: Training-Hardware Mismatch}}

The fundamental obstacle to LUT-native learning lies in the mathematical discrepancy between the discrete Boolean nature of FPGA LUTs and the continuous gradient-based optimization. Drawing from established principles in BNN research~\cite{courbariaux2015binaryconnect,liu2018bi}, optimizing within a continuous domain typically yields higher training stability and performance compared to direct discrete domain training. While the logic of a LUT is distinct from the bit-wise arithmetic in BNNs, a LUT also operates within the binary domain and thus follows similar principles and necessitates a method to bridge discrete and continuous representations. 

Existing differentiable LUT formulations do not satisfy gradient consistency while remaining scalable. LUTNet~\cite{wang2020lutnet} employs Lagrange interpolation for relaxation but fails to align with its BNN architecture. DLN and LTN~\cite{petersen2022deep,petersen2024convolutional} assign a trainable weight to every single candidate in the enumerated Boolean function space. As a $k$-input LUT contains $2^k$ bits which can represent $2^{2^k}$ distinct Boolean functions, the parameter complexity scales double-exponentially. For the widely used 6-LUT in modern FPGAs~\cite{boutros2021fpga}, this requires searching across $2^{2^6} = 2^{64} \approx 1.8\times10^{19}$ functional weights per LUT, which is prohibitively expensive and fundamentally unscalable. DWN~\cite{bacellar2024differentiable} attempts to enhance learning by augmenting the precise discrete gradient function~\cite{strikwerda2004finite} of LUT logic with an extension term. This heuristic introduces a mismatch between forward and backward that undermines optimization consistency. These limitations prevent LUT-native networks from scaling to complex tasks while maintaining network performance. This motivates us to develop a differentiable LUT formulation that remains gradient-consistent, scales efficiently to 6-LUTs, and recovers hardware logic at its discrete boundaries.

\noindent\textbf{\textit{Challenge 2: Topology-Architecture Co-Design under FPGA Physical Constraints}}

While individual LUTs are expressive, practical deployment requires organizing them into a network topology. A fundamental gap exists between the unconstrained algorithmic topology and the rigid routing fabric of FPGA physical architectures. Existing LUT-as-neuron methods often adopt unstructured (random or learnable) inter-neuron connectivity to maximize network performance. Our analysis reveals that such unstructured topologies are physically incompatible as they lead to non-local routing, forcing signals to traverse long distances across the FPGA fabric.

\begin{figure}
    \centering
    \includegraphics[width=0.8\linewidth]{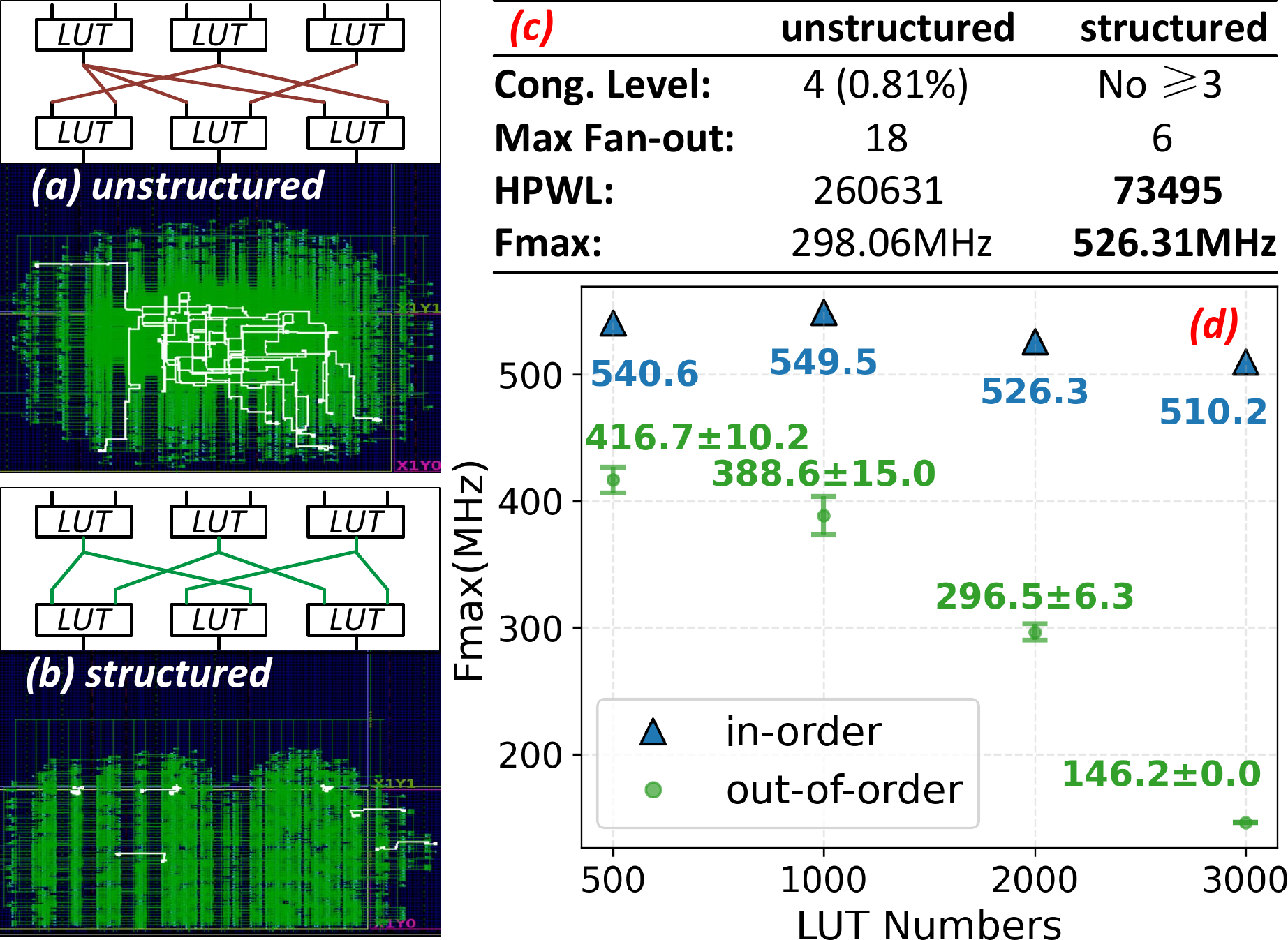}
    \caption{Profiling of connection topologies. (a, b) Post-P\&R layout of unstructured and structured designs, highlighting top 10 critical paths. (c) Quantitative P\&R results. The structured topology drastically reduces wirelength and achieve higher frequency. (d) Scaling behavior of $F_{\text{max}}$ versus LUT count. Structured topology maintains high frequency while the unstructured one causes severe frequency degradation. }
    \label{fig_order}
\end{figure}

To quantify this, we profile an MLP-style LUT-native network based on DWN~\cite{bacellar2024differentiable}, the SOTA 6-LUT based work, consisting of two 6-LUT layers with 2000 neurons each. Comparing unstructured (random) and structured connections based on post-placement and routing (post-P\&R) implementation results, we observe that the former produces convoluted timing paths (Fig.~\ref{fig_order}(a)) while the latter yields a much more localized routing (Fig.~\ref{fig_order}(b)). Quantitatively, the structured topology achieves a $3.5\times$ reduction in half-perimeter wirelength (HPWL) (Fig.~\ref{fig_order}(c)). This reduction enables the structured design to reach 526MHz, whereas the unstructured topology is restricted to 298MHz. Furthermore, as the number of LUTs per layer increases, the structured topology maintains a high and relatively stable maximum clock frequency ($F_{\text{max}}$), while the unstructured alternative shows significant frequency degradation (Fig.~\ref{fig_order}(d)). These results indicate that unstructured connectivity can directly undermine the physical implementability of LUT-native networks.

Beyond restricting connectivity, this lack of locality also stems from the neglect of well-studied neural network principles, such as the hierarchical aggregation and local receptive fields inherent to CNN-style topologies. These principles are also naturally compatible with the FPGA's localized fabric. Moreover, a dedicated topology alone is insufficient. Even with structured local connectivity, a naive mapping of a LUT-native CNN to hardware can result in excessive combinational logic depth, inter-layer throughput mismatches, and inefficient data movement. Addressing these bottlenecks requires a tight co-design of structured topology and a streaming architecture that preserves locality, balances dataflow, and ensures high-frequency timing closure.

\noindent\textbf{\textit{Challenge 3: Absence of Systematic Design Space Exploration}}

The realization of a LUT-native network fundamentally redefines the accelerator's design space, yet there remains an absence of systematic methodologies to explore its unique performance-resource trade-off. In conventional FPGA-based CNN accelerators, DSP and memory blocks are typically the dominant bottlenecks, while LUTs are treated as inexhaustible auxiliary glue logic~\cite{zhang2015optimizing,liu2017throughput}. However, the LUT-as-neuron paradigm creates a DSP/memory-free architecture, shifting the key design bottleneck to LUT utilization coupled with pipeline depth and inter-layer throughput coordination. 

This fundamental shift renders conventional DSP-centric analytical QoR models~\cite{ma2019performance, huang2021fpga} inapplicable, while prior LUT-native works~\cite{wang2020lutnet,petersen2024convolutional} remain restricted to hand-crafted implementations that fail to explore the broader design landscape. This absence of systematic exploration necessitates high-fidelity LUT-centric analytical QoR model and a latency-driven compilation framework for LUT-native streaming accelerators. Such a framework should automatically explore the performance-resource trade-off and generate optimized FPGA implementations, bridging the gap from well-trained networks to hardware realizations.

\subsection{Our Solution}
To address the above challenges, we develop FPGN, a LUT-native framework for nanosecond-scale neural accelerators on FPGA, with jointly designed differentiable training, physically-aware topology and architecture design, as well as automated compilation. 
\begin{itemize}
    \item First, to enable training-hardware consistency for LUT-native networks on FPGA, we introduce a differentiable relaxation for LUTs together with a progressive binarization strategy that gradually transitions the model toward exact binary deployment. 
    \item Leveraging the trainable LUT neurons, we design a physically-aware LUT-native topology with structured connectivity and a fully streaming hardware architecture to enable high-frequency neural acceleration by efficiently truncating long combinational paths.
    \item Finally, to bridge trained networks and actual FPGA implementations, we develop a latency-driven compiler equipped with high-fidelity LUT-centric analytical QoR models, which guides design space exploration~(DSE) and automatically generates efficient RTL implementations. 
\end{itemize}

Together, they establish a unified path from algorithmic training to nanosecond-scale neural inference on FPGA.

\section{Hardware-Aligned Training Methodology}\label{sec:training}
We first address the training-hardware mismatch challenge by developing an training method for LUT-native networks that remains consistent with exact FPGA hardware. The key idea is to formulate LUT computation in a differentiable relaxation and progressively drive this relaxation toward discrete boundaries, which enables efficient gradient-based training of LUT-native networks built on modern 6-LUT primitives. For simplicity, we refer to LUT configuration bits as \textit{weights} throughout the remainder of this section.

\subsection{Differentiable LUT Formulation}\label{subsec:diffLUT}

The core challenge in training LUT-native networks is that the LUT primitive is fundamentally discrete, which prevents direct gradient-based optimization. To bridge this gap, we relax the discrete LUT by explicitly modeling its hardware decoding logic via discrete equality indicator and introduce differentiable equality indicator.

A hardware $k$-LUT is defined by $k$ input bits $\mathbf{x}\in\{0,1\}^k$ and $2^k$ configuration bits $\mathbf{w}\in\{0,1\}^{2^k}$. Its output is selected from $\mathbf{w}$ by the discrete address encoded by $\mathbf{x}$:
\begin{equation}\label{eq_lut}
    \operatorname{idx}(\mathbf{x})=\sum_{i=0}^{k-1}x_i\,2^i,\quad
    f_k(\mathbf{x},\mathbf{w})=\sum_{\mathbf{u}\in\{0,1\}^{k}}w_{\operatorname{idx}(\mathbf{u})}\cdot \mathbb{I}(\mathbf{x}; \mathbf{u})
\end{equation}
where $\operatorname{idx}(\cdot)$ converts a binary vector to an integer look-up address, and $\mathbb{I}(\mathbf{x};\mathbf{u})$ is a discrete equality indicator to select the specific entry in the LUT that evaluates to 1 if the input $\mathbf{x}$ exactly matches the specific address pattern $\mathbf{u}$, and 0 otherwise. In digital logic, the selection process is implemented via an address decoder. For each bit position $i$, the circuit checks for a match using the Boolean product: $x_i$ if $u_i = 1$, and $\bar{x}_i$ if $u_i = 0$. Therefore, within the strict binary domain $\mathbf{x} \in \{0,1\}^k$, the decoder can be explicitly written as the following product:
\begin{equation}\label{eq_discrete_product}
    \mathbb{I}(\mathbf{x}, \mathbf{u}) = \prod_{i=0}^{k-1} x_i^{\,u_i}\,(1-x_i)^{1-u_i}=
    \begin{cases}
        1 & \text{if } \mathbf{u} = \mathbf{x} \\
        0 & \text{otherwise}
    \end{cases}
\end{equation}

To make it trainable, we relax the logic selection into a continuous process by allowing input to take real values, i.e., $\mathbf{x}\in \mathbb{R}^k$.

\noindent{\textbf{Differentiable Equality Indicator:}} For a given vector $\mathbf{x}\in \mathbb{R}^k$ and a binary pattern $\mathbf{u}=(u_{k-1},\dots,u_0) \in \{0,1\}^k$, the equality indicator is defined as:
\begin{equation}\label{eq_equality_indicator}
    \delta(\mathbf{x};\mathbf{u}) =\prod_{i=0}^{k-1} x_i^{\,u_i}\,(1-x_i)^{1-u_i}
\end{equation}

This function equals 1 when $\mathbf{x}=\mathbf{u}$ and smoothly decreases to 0 at the bitwise complement $\bar{\mathbf{u}}$. With this, the LUT output can be relaxed to the following continuous function:
\begin{equation}\label{eq_float_lut}
    f_k(\mathbf{x},\mathbf{w})=\sum_{\mathbf{u}\in\{0,1\}^k} w_{\operatorname{idx}(\mathbf{u})}\, \prod_{i=0}^{k-1} x_i^{\,u_i}\,(1-x_i)^{\,1-u_i}
\end{equation}

With this differentiable formulation, the output becomes a linear combination of all weights, allowing gradients to flow to every weight and input during backpropagation. When the inputs $\mathbf{x}$ are restricted to the binary domain $\{0,1\}^k$, the relaxed formulation reduces exactly to the discrete hardware LUT lookup, as only one $\delta(\mathbf{x};\mathbf{u})$ equals one. At these discrete boundaries, the gradient with respect to the LUT weights becomes:
\begin{equation}\label{eq_sparse_grad}
    \frac{\partial f_k(\mathbf{x}, \mathbf{w})}{\partial w_{\operatorname{idx}(\mathbf{u})}} =
    \begin{cases}
        1 & \text{if } \mathbf{u} = \mathbf{x} \\
        0 & \text{otherwise}
    \end{cases}
\end{equation}

This proves that our formulation exactly recovers discrete LUT logic at binary inputs while enabling gradient-based optimization, thereby addressing the gradient mismatch problem of EFD in DWN~\cite{bacellar2024differentiable}. Additionally, this formulation provides a scalable computational foundation for training high-input LUT primitives. While the formulation exhibits $\mathcal{O}(2^k)$ complexity per neuron, it represents the minimal parameterization required to span the entire $2^{2^k}$ Boolean functional space. Unlike prior logic-native works~\cite{petersen2022deep, petersen2024convolutional} that suffer from double-exponential parameter explosion ($\mathcal{O}(2^{2^k})$), FPGN aligns the optimization complexity with the physical capacity of LUT primitives. For the 6-LUTs prevalent in modern FPGAs, this reduces the number of trainable parameters per neuron from $2^{2^6}$ function-selection weights to 64 LUT entries, addressing the scalability problem in Challenge 1.

\begin{figure}
    \centering
    \includegraphics[width=0.9\linewidth]{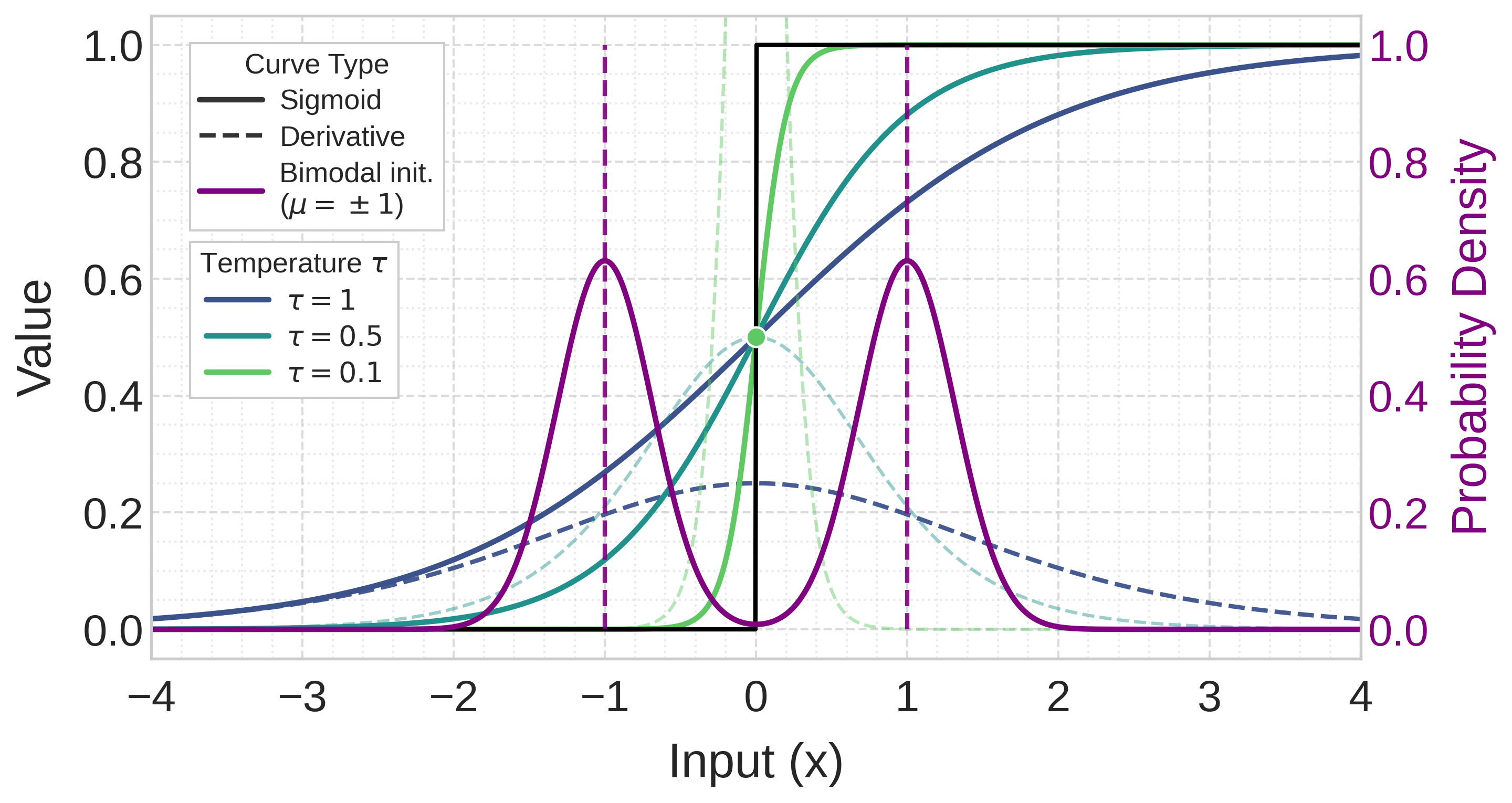}
    \caption{Sigmoid and its derivative for different temperatures \(\tau\), together with the bimodal Gaussian initialization centered at \(\mu=\pm1\). Smaller \(\tau\) yields a sharper approximation to binary thresholding and a narrower informative-gradient region. The initialization at \(\pm1\) is used only at the start of training (\(\tau=1\)) and serves as an empirical compromise between weak multiplicative gradients and sigmoid saturation.}
    \label{fig_sigmoid_and_deriv}
\end{figure}

\subsection{Bridging the Continuous-Discrete Gap}

While the differentiable formulation enables gradient-based optimization, directly training LUTs with discrete binary weights $\mathbf{w} \in \{0,1\}^{2^k}$ as DWN~\cite{bacellar2024differentiable} remains highly unstable~\cite{courbariaux2015binaryconnect,liu2018bi} and inefficient due to extreme gradient sparsity. As Eq.~\eqref{eq_sparse_grad} shows, only the single weight indexed by current input $\mathbf{x}$ receives an update in each iteration, while the gradients for the remaining $2^k-1$ weights vanish. This extreme gradient sparsity fundamentally hinders the effective exploration of the massive Boolean functional space.

To overcome this, we adopt a three-stage optimization strategy: (i) stable training on smooth continuous relaxation, (ii) progressive binarization to bridge the quantization gap, and (iii) binary fine-tuning for hardware alignment. Inspired by BNN works~\cite{liu2018bi, qin2020forward}, we map real-valued weights to soft binary values through a sigmoid function with an adjustable temperature $\tau$:
\begin{equation}\label{eq_gumbel_sigmoid}
        w_q = \sigma(w)=\operatorname{sigmoid}(\frac{w}{\tau}) = \frac{1}{1+e^{-\frac{w}{\tau}}}
\end{equation}

As shown in Fig.~\ref{fig_sigmoid_and_deriv}, a larger $\tau$ yields a smoother relaxation, while a smaller $\tau$ makes the mapping increasingly close to hard binary thresholding, pushing more relaxed weights toward the binary extremes. We therefore adopt a progressive annealing schedule for $\tau$ so that training starts from a smooth continuous relaxation and gradually approaches discrete binarization. This continuous stage mainly serves as an optimization mechanism that makes the gradient flow more dense to improve training efficiency. After this relaxed stage, training is further continued in the binary domain based on the gradient in Eq.~\eqref{eq_sparse_grad}, so that the learned LUTs are ultimately aligned with the deployed Boolean form. 

\textbf{Gradient Attenuation Analysis:} While soft binarization enables continuous optimization, gradients can attenuate severely during training due to two issues. First, the repeated multiplications in the differentiable equality indicator (Eq.~\eqref{eq_equality_indicator}) naturally shrink gradients as they propagate to LUT inputs. Second, the sigmoid used for soft binarization function narrows the effective gradient region as $|w|$ increases as shown in Fig.~\ref{fig_sigmoid_and_deriv}. To mitigate these effects, we initialize weights via a bimodal distribution using Gaussian distributions. Since this initialization is applied only at the beginning of training, it is interpreted with respect to the \(\tau=1\) sigmoid curve in Fig.~\ref{fig_sigmoid_and_deriv}. We set the centers of the bimodal distribution to \(\pm1\) empirically as a practical compromise. Weights of this scale are sufficiently separated from zero to avoid overly weak multiplicative gradients in the LUT relaxation, while still remaining outside the strongly saturated region of the sigmoid. This initialization therefore allows gradients to flow across layers while preserving informative updates during the continuous optimization stage.

By combining progressive annealing with bimodal initialization, this three-stage optimization strategy maintains stable gradient propagation for efficient training, while ensuring bit-precise alignment with the final FPGA implementation. Together with the scalable differentiable formulation, the hardware-aligned training methodology provides a robust solution to the inherent training-hardware mismatch.

\section{Physically-Aware Network Topology and Hardware Architecture}\label{sec:architecture}
Building on our training foundation, we now scale individual LUT primitives into physically-aware network topology and their physical implementations on FPGAs. To bridge the gap between unconstrained algorithmic connectivity and the rigid FPGA routing fabric (Challenge 2), we propose a hierarchical co-design framework spanning three levels: micro-topology, macro-topology, and streaming architecture. Our core strategy is to enforce structural regularity throughout the hierarchy and transform abstract neural connections to deterministic FPGA dataflow.

\subsection{Structured LUT-Native Micro-Topology}
To bridge the gap between the limited fan-in of a single $k$-LUT and the coordinated operations across high-dimensional feature spaces of neural networks, we define structured micro-topologies following the well-studied functionality from traditional CNNs, i.e., parallel processing and reduction units.

\textbf{Parallel Processing Units:} To process input streams in parallel, we design the LUT-vector as the basic computation unit. A LUT-vector with $N_o$ output bits is represented as a mapping $I\in\{0,1\}^{k\times N_o}\to O\in\{0,1\}^{N_o}$, where each output bit is generated by an independent $k$-LUT. As illustrated in Fig.~\ref{fig_architecture}(c), LUT-vector can be used to construct layers analogous to fully-connected~(FC) layers where a $k$-LUT is a neuron.

\textbf{Reduction Units:} We distinguish two reduction types based on their output domains. \textbf{Binary Reduction:} To aggregate multiple input bits into a single binary activation ($I\in\{0,1\}^{N_i}\to O\in\{0,1\}$), we employ a LUT-tree. As shown in Fig.~\ref{fig_architecture}(d), it forms a decreasing pyramidal structure of $\lceil \log_k N_i \rceil$ cascading LUT-vectors. This logarithmic depth ensures that the combinational delay increases minimally as input width scales, sustaining high-frequency operation. \textbf{Integer Reduction:} For scenarios that need sufficient precision, we utilize popcount units for binary-to-integer transitions ($I\in\{0,1\}^{N_i}\to O\in\mathbb{Z}$).

\textbf{In-Order Flattening:} After algorithmically designing these units, the structured interconnection between these topologies is critical for physical efficiency. In traditional arithmetic-centric neural networks, data ordering is merely a memory indexing convention. However, in our LUT-native streaming architecture, the algorithmic data order is strictly equivalent to the physical hardware interconnection and pin assignment. By in-order flattening high-dimensional data (e.g., feature maps) into bit-vectors (Fig.~\ref{fig_architecture}(a)), we ensure that the output bits of one stage are fed into the next stage with local routing. 

\textbf{Locality-Aware Padding:} To maintain this structured connectivity when data dimensions across adjacent layers are mismatched, we propose locality-aware padding that processes the data sequence through a two-stage padding mechanism. Suppose a LUT-vector comprises $N_{\text{total}}$ LUTs while the size of input bit-vector is $M<k\times N_{\text{total}}$. First, we allocate the unique input bits to the initial $N_{\text{base}}=\lceil M/k \rceil$ LUT units. If the $N_{\text{base}}$-th LUT lacks sufficient inputs, we perform a local padding to get $\hat M=N_{\text{base}}\times k$ bits that vacant inputs are filled by sequentially repeating the available bits within the same LUT. Second, for the remaining $(N_{\text{total}}-N_{\text{base}})$ LUTs, we replicate the $\hat{M}$ bits in order until all LUTs are fully occupied. Algorithmically, this dual-mechanism approach increases the average information utilization per LUT by ensuring no logic gates are left idle. From a hardware perspective, the first stage guarantees local connectivity while the second stage effectively distributes the fan-out pressure across all input bits rather than overloading a single source, preventing individual signal nets from becoming timing bottlenecks and ensuring a high operating frequency ($F_{\text{max}}$).

In essence, these structured micro-topologies are interconnected via in-order connections with locality-aware padding, ensuring routing locality and mitigating long-distance routing wires to sustain high operating frequency. This provides the foundation that enables us to build a sophisticated global network topology capable of complex tasks.

\subsection{Hardware-Aware Network Topology Design}~\label{sec_4_topology}
\begin{figure}
    \centering
    \includegraphics[width=0.9\linewidth]{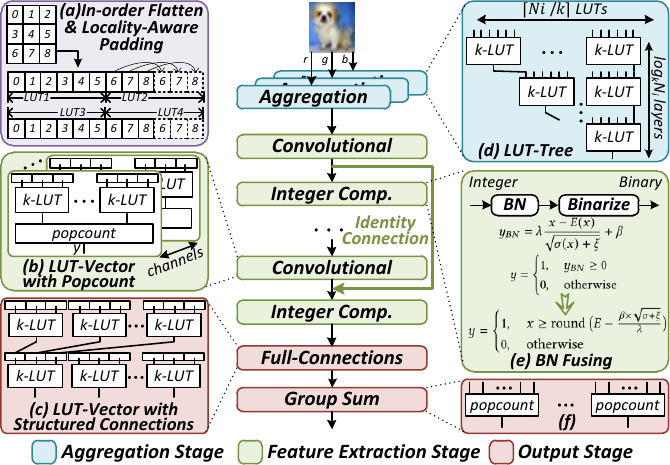}
    \caption{FPGN's network topology. The network consists of hardware-friendly layers built on structured micro-topology (a-d,f). (e) Mathematical fusion of BN and binarization into a simple integer comparison.}
    \label{fig_architecture}
\end{figure}

While micro-topologies ensure local efficiency, cascading them blindly across a deep network risks global routing congestion and precision mismatch. To preserve this local efficiency macroscopically while obtaining high algorithmic accuracy, we compose these primitives into a hierarchical, three-stage topology as shown in Fig.~\ref{fig_architecture} that mirrors traditional CNN structural paradigms to sustain high representation capacity on the FPGA fabric.

\textbf{Aggregation Stage:} To mitigate information loss and eliminate costly floating-point preprocessing, we replace conventional quantization with a learned, pixel-wise aggregation layer. It is a convolutional-style layer featuring a LUT-tree as the $1\times 1$ convolutional kernel to aggregate a pixel's spatial bit-vector, representing different input channels into a single informative activation. By allowing multiple aggregation channels, this mechanism not only aggregates the full input information into each bit but also enables diverse feature extraction across output channels using purely LUT-native operations.

\textbf{Feature Extraction Stage:} This stage comprises a stack of residual blocks with LUT-native convolutional (LUT-Conv) layers implemented as channel-wise parallel LUT-vectors followed by popcount units to ensure sufficient precision for the subsequent layers. By performing an inherent $k$-to-1 mapping within each LUT, this architecture achieves a $k$-fold complexity reduction to popcount units, shrinking the adder-tree depth from $\mathcal{O}(\log(N))$ to $\mathcal{O}(\log(N/k))$. To sustain high-frequency timing closure, we maintain architectural homogeneity by eliminating floating-point routing overheads. Inspired by~\cite{he2016identity}, we reposition the identity addition before the BN layer, ensuring that residual additions are performed directly on low-bit integer outputs from popcounts. Consequently, BN layers can be mathematically fused into the subsequent non-linear binarization function as a simple integer comparison during inference (Fig.~\ref{fig_architecture}(e)), allowing all components to be mapped onto LUTs to completely eliminate DSP usage from our design.

\textbf{Output Stage:} The final stage comprises cascaded FC-layers (Fig.~\ref{fig_architecture}(c)) and a task-specific unit. For classification, the task-specific unit implements a popcount-based group-sum operation to produce the final scores with high numerical precision required for robust classification performance~\cite{umuroglu2017finn, rastegari2016xnor}, as illustrated in Fig.~\ref{fig_architecture}(f). 


By connecting these macro blocks through structured connections to maintain the locality of their underlying micro-topologies, we establish a hardware-aware network topology that eliminates floating-point dependencies and reduces arithmetic complexity. This deterministic, LUT-native representation provides the structural regularity required for the fully streaming dataflow.

\begin{figure*}
    \centering
    \includegraphics[width=1.8\columnwidth]{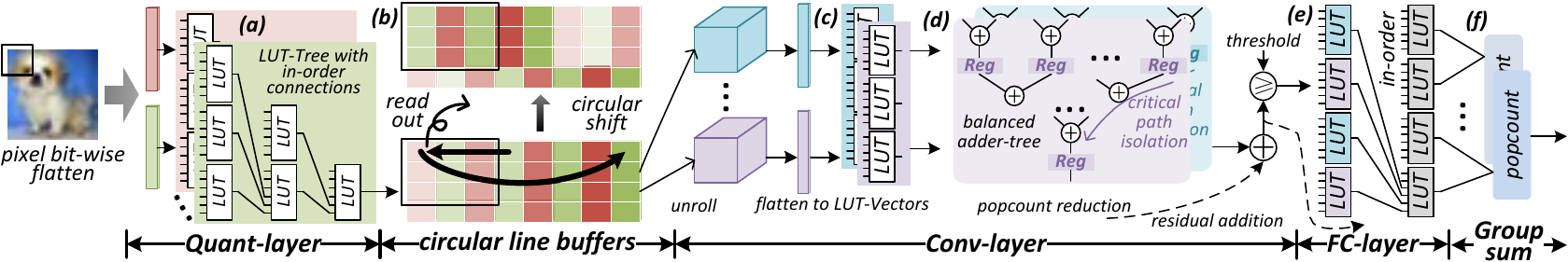}
    \caption{Overview of the FPGN streaming dataflow architecture. Inter-layer circular buffers decouple adjacent layers to sustain concurrency. At the intra-layer level, pipeline registers are adaptively inserted to break long timing path.}
    \label{fig_hardware}
\end{figure*}

\subsection{Fully Streaming Hardware Architecture}
To implement the hardware-aware topology as a physical accelerator to achieve nanosecond-scale neural acceleration, we propose a fully streaming architecture as illustrated in Fig.~\ref{fig_hardware}, to optimize the LUT footprint for various FPGA resource constraints while simultaneously breaking down long combinational paths into a pipelined execution flow.

\textbf{Flexible Intra-Layer Spatial Computation:} To dynamically adjust LUT resource utilization under different resource constraints, we design the processing units of convolutional-style layers (the aggregation and LUT-Conv stages) to support configurable unrolling by leveraging the kernel-sharing property of convolution~\cite{ma2019performance}. By scaling the unroll factor of these kernels, the architecture allows for an adaptive trade-off between diverse hardware resource budgets and inference latency. 

\textbf{Adaptive Intra-layer Pipelining:} To ensure high-frequency timing closure, we implement an adaptive pipelining strategy within the processing units. Intra-layer processing units are based on two primitive types, LUT-vectors (Fig.~\ref{fig_hardware}(a, c, e)) and popcount reduction units comprising balanced adder trees (Fig.~\ref{fig_hardware}(d, f)). Based on the logic depth of cascaded LUTs, pipeline registers are strategically inserted to maintain consistent timing. Furthermore, since the integer adders in popcount units are implemented to utilize FPGA Carry-Chain primitives~\cite{ug474}, we adaptively insert registers based on the accumulated number of Carry-Chain stages to eliminate combinational bottlenecks.

\begin{table}[tbp]
    \centering
    \caption{Symbol Definitions for the Analytical QoR Model}
    \label{tab_symbol}
    \begin{tabular}{c l}
        \toprule
        \textbf{Symbol} & \textbf{Description} \\
        \midrule
        \multicolumn{2}{l}{\textit{\textbf{Network Layer Parameters (Inputs)}}} \\
        $W_i, H_i, C_i^{\text{in}}, C_i^{\text{out}}$ & Width, height, and channels. \\
        $k_i, s_i$ & Kernel size, stride. \\
        \midrule
        \multicolumn{2}{l}{\textit{\textbf{Design Parameters (Inputs)}}} \\
        $w_i, h_i$ & Column-wise and row-wise unroll factors. \\
        \midrule
        \multicolumn{2}{l}{\textit{\textbf{Resource Model (Outputs \& Variables)}}} \\
        $N_{l,i}, N_{r,i}$ & Resource consumption of LUT and Reg. \\
        $B_i$ & Number of input bits to a processing unit. \\
        \midrule
        \multicolumn{2}{l}{\textit{\textbf{Latency Model (Outputs \& Variables)}}} \\
        $St_i,Pt_i$ & Start cycle \& Processing time.\\
        $T_i^{chunk}$ & Processing time of a chunk.\\
        $T_i^{\text{startup}}$ & The time waiting for the startup data.\\
        $It_i$, $D_i$ & Pipeline interval and delay.\\
        $L_{\text{total}}$ & Total latency of the entire network inference. \\
        \bottomrule
    \end{tabular}
\end{table}

\textbf{LUT-Efficient Inter-Layer Streaming:} To support the unrolling mechanism and sustain high-throughput streaming, we propose a stationary-window circular line buffer (Fig.~\ref{fig_hardware}(b)) for inter-layer decoupling. These buffers enable concurrent write and read operations, ensuring a continuous dataflow. Standard unrolling typically results in complex, multi-address access patterns, necessitating LUT-consuming and timing-critical multiplexers~(MUXs) to select data from various buffer offsets based on sliding window indices. Our design mitigates this by fixing the read window and shifting the data instead, effectively replacing high-fan-in selection logic with deterministic structural interconnects. Specifically, the buffer performs horizontal circular shifting during row-wise streaming and vertical row-promotion upon the completion of each line. Since these shifting patterns are hard-wired post-synthesis for a given unroll factor, the fan-in for each register remains minimal and constant. This optimization strategically reduces MUX complexity, thereby minimizing the total LUT footprint, ensuring high-frequency operation and routing feasibility.

Based on these dedicated LUT and timing-oriented hardware optimizations, the fully streaming hardware architecture achieves a deterministic balance between high-throughput execution and high-frequency timing closure. By organizing individual LUT neurons into structured topology and mapping it to the LUT-native hardware design, we successfully translate the algorithmic advancement of differentiable LUTs into a high-performance FPGA realization. This architecture provides a flexible hardware template and defines the design space, enabling our compiler to systematically automate spatial unrolling and performance optimization across diverse resource-constrained scenarios.

\section{Latency-Driven Automated Compiler}
While the streaming architecture provides a high-performance template, the massive design space renders manual optimization impractical, comprising diverse unrolling factors across multiple layers. To bridge this gap, we propose a latency-driven automated compiler that systematically navigates the trade-offs between hardware concurrency and resource constraints. To achieve this, we first establish an analytical QoR model to evaluate latency and resource consumption with high-fidelity by leveraging the deterministic nature of our LUT-native architecture. Based on this  mathematical foundation, the compiler employs a hybrid DSE strategy combining heuristic search with Mixed-Integer Linear Programming~(MILP) solver to search for the optimal configurations. Finally, the compiler instantiates the optimized parameters into a synthesizable RTL design using a library of pre-defined hardware templates.

\subsection{High-Fidelity Analytical QoR Modeling}
To enable precise DSE without the need for time-consuming logic synthesis, we develop an analytical QoR model that translates algorithmic parameters into physical QoR metrics by exploiting the deterministic mapping between our hardware primitives and the FPGA fabric. The requisite symbols are defined in Table~\ref{tab_symbol} and key analytical formulations are summarized in Table~\ref{tab_performance_modeling}.

\subsubsection{Timing Modeling}
To ensure the generated architecture meets the target clock frequency ($F_{\text{max}}$), the timing model quantifies combinational delays based on the specific physical implementation of two components: the logic depth of cascaded LUTs and the stages of Carry-Chain based adders~\cite{ug474}. By tracking these propagation delays, the compiler determines the optimal intervals for pipeline register insertion. This process truncates long combinational paths and ensures that the critical path delay remains within the system clock period.

\begin{table}[tbp]
\caption{Key Equations for the Analytical QoR Model}
\label{tab_performance_modeling}
\centering
\resizebox{\columnwidth}{!}{%
\begin{tabular}{c c l}
\toprule
\textbf{Aspect}      & \textbf{Component}        & \textbf{Equation} \\
\midrule
\multicolumn{3}{l}{\textit{\textbf{Resource Model}}} \\
\multirow{5}{*}{LUT} & \multirow{4}{*}{Total} & 
    $N_{l,i}^\text{Quant}  = N_{l,i}^{\text{kernel}} + N_{l,i}^{\text{mux}}$ \\
    &&$N_{l,i}^\text{Conv} = N_{l,i}^{\text{kernel}} + N_{l,i}^{\text{arith}} + N_{l,i}^{\text{mux}}$  \\
    &&$N_{l,i}^\text{Out} = \sum (\lceil B_i/k_i\rceil + N_{l,i}^{\text{grp\_sum}})$ \\ \addlinespace
    & Kernel & $N_{l,i}^{\text{kernel}} = (\lceil B_i/k_i\rceil \cdot C_i^{\text{out}}) \cdot w_i \cdot h_i$\\
\midrule
\multirow{3}{*}{Reg} & Total & 
    $N_{r,i}^\text{Conv/Agg} = N_{r,i}^{\text{buffer}}(h_i, h_{i+1}, k_{i+1}, ...)+ N_{r,i}^{\text{pipe}}$ \\ \addlinespace
    & Chunk  & $N_{r,i}^{\text{chunk}}=(k_i+(w_i-1)\times s_i)\times C_i^{\text{out}}\times W_i$\\ \addlinespace
    & Buffer & $N_{r,i}^{\text{buffer}} = N_{r,i}^{\text{chunk}} + \max(N_{r,i}^{\text{consumer}}, N_{r,i}^{\text{producer}})$ \\
\midrule
\multicolumn{3}{l}{\textit{\textbf{Latency Model}}} \\
\multicolumn{2}{l}{Processing Time} & $Pt_i = (H_i/h_i - 1) \cdot It_i + T^{\text{chunk}}_i$ \\
\multicolumn{2}{l}{Start Time} & $St_i = St_{i-1}+T_i^{\text{startup}} (+ D_{i-1}\quad\text{if not Agg layer}) $\\
\bottomrule
\end{tabular}%
}
\end{table}

\subsubsection{Resource Utilization Modeling}

Since the proposed streaming architecture is purely LUT-native, it eliminates dependencies on heterogeneous blocks such as DSPs or BRAMs. Consequently, the resource model focuses exclusively on LUTs ($N_{l,i}$) and registers ($N_{r,i}$), where the total consumption is defined as the analytical summation across all layers based on the deterministic structural properties of the hardware. The LUT utilization ($N_{l,i}$) for layer $i$ is defined by the logic required for its LUT-native units, arithmetic logic, and MUXs of circular line buffers, while registers are mainly utilized for inter-layer line buffers and pipelining.

\textbf{LUT-native Kernels ($N_{l,i}^{\text{kernel}}$):} The model accounts for the direct mapping of processing units to $k$-LUT primitives. The LUT count is determined by the input bit-width ($B_i$) and the output channels ($C_i^{\text{out}}$). Under the unrolling mechanism, these costs are scaled by the unroll factors ($w_i \cdot h_i$) to reflect the parallel instantiation of the hardware units. This relationship ensures that the LUT footprint is predictable given the degree of spatial parallelism.

\textbf{Arithmetic Logic ($N_{l,i}^{\text{arith}}$):} The arithmetic logic includes popcount units, as well as other integer operation such as adders and comparisons within residual blocks. Since a popcount operation comprises a tree of integer adders, and both adders and comparators are implemented using Carry-Chain primitives, their LUT costs is modeled as a function of the operand bit-width. Specifically, the total popcount cost ($N_{l,i}^{\text{popcount}}$) is a function of the compressed bit-width ($B_i/k$), reflecting the $k$-fold complexity reduction introduced in Sec.~\ref{sec_4_topology}. Similarly to LUT-native kernels, this cost is also scaled by layer-specific unroll factors to ensure consistency with the overall datapath parallelism.

\textbf{Buffering MUXs ($N_{l,i}^{\text{MUX}}$):} The LUT overhead for data shifting is modeled based on the MUX requirements of the circular line buffers. According to the stationary read window design and the shift mechanism, the fan-in for each buffer register remains constant for a given set of unroll factors. This allows the compiler to calculate the LUT resources requirement based on the fixed number of input sources to each MUX~\cite{ug474}.

\textbf{Register Modeling ($N_{r,i}$):} Register consumption is divided into inter-layer line buffer cost ($N_{r,i}^{\text{buffer}}$) and pipelining cost ($N_{r,i}^{\text{pipe}}$) for maintaining high-throughput streaming. In our design, since each channel is physically instantiated, the line buffer supports all channels simultaneously, making its size ($N^{\text{buffer}}_{r,i}$) a deterministic function of the number of rows. To quantify the buffering cost, we define a \textit{chunk} as the minimum set of input rows required to sustain the sliding-window computation, whose size is determined by the row-wise unroll factor ($h_i$) and network configurations (kernel size $k_i$, and stride $s_i$). To ensure unblocked inter-layer streaming, the line buffer is designed to satisfy the requirements of both the consumer and the producer. Specifically, the buffer is sized to store at least one active chunk for the current processing task while providing additional rows to meet both the subsequent chunk requirement by the consumer layer and new data bursts requirement from the producer layer. Additionally, the pipeline register count ($N_{r,i}^{\text{pipe}}$) is derived from the structural timing model to break combinational logic paths for satisfying timing requirement. 

\subsubsection{Streaming Latency Modeling}

The end-to-end latency ($L_{\text{total}}$) is defined as the completion time of the final layer in the streaming architecture, which is determined by a layer-wise model of two fundamental components: start time ($St_i$) and processing time ($Pt_i$).

\textbf{Processing time ($Pt_i$):} This represents the steady-state execution time to process all data. Inspired by a Roofline-analysis \cite{zhang2015optimizing}, it is constrained by (i) the latency for processing a single data chunk ($T_i^{\text{chunk}}$)~(compute-bound), and (ii) the pipeline interval ($It_i$), the time between launching two consecutive chunks~(similar to memory-bound).

\textbf{Start time ($St_i$):} The start time captures the initial starvation period before a layer begins its first processing. For the first layer, $St_i$ is determined by the system bandwidth ($BW$) required to fill the initial buffer. For subsequent layers, $St_i$ represents the cumulative time required for the producer to generate a valid data chunk, plus the pipeline delay ($D_i$).

By integrating these timing, resource, and latency models, we establish a comprehensive analytical QoR model that captures the complex interactions between algorithmic unrolling and physical hardware constraints, transforming the streaming hardware architecture into a mathematical foundation for the automated DSE.

\subsection{Latency-Driven Design Space Exploration}
With the analytical QoR model established, the goal of DSE is to identify the optimal set of unroll factors $\{w_i,h_i\}$ for each layer $i$ that minimizes the total inference latency ($L_{\text{total}}$) under strict FPGA resource and system bandwidth constraints. This optimization problem is inherently non-linear and non-convex, primarily due to the max-based dependencies in pipeline intervals and cascaded products of unroll factors across layers introduced by the start-time ($St_i$) dependence. To solve this optimization efficiently, we decompose the exploration into a two-level hybrid strategy that combines a heuristic search with Mixed-Integer Linear Programming (MILP).

\textbf{Outer-Level Heuristic Search.} We utilize a steepest-descent search to explore the row-wise factors $\{h_i\}$. Since $\{h_i\}$ directly governs the line buffer capacity and the inter-layer synchronization delay ($St_i$), it serves as the primary source of non-linearity. By isolating $\{h_i\}$ in the outer loop, the remaining sub-problem can be transformed into linear formulation.

\begin{table}[tbp]
\caption{Experimental Setup}
\label{tab_setup_compact}
\centering
\begin{threeparttable}
\small
    \begin{tabularx}{\columnwidth}{@{}l l >{\raggedright\arraybackslash}X@{}}
    \toprule
    \textbf{Model} & \textbf{Layer (Reps)} & \textbf{Configuration} \\
    \midrule
    FPGN-3   & Agg ($\times$1) & $C_{\text{in}}=8$, $C_{\text{out}}=(m^{*}//3)\cdot k^\dagger$ \\
             & Conv ($\times$3)  & Stride=2, $C_{\text{out}}=m^{*}\cdot k^\dagger$ \\
             & FC ($\times$2)    & 2000 LUTs per layer \\
    \midrule
    FPGN-6   & Agg ($\times$1) & $C_{\text{in}}=8$, $C_{\text{out}}=(m^{*}//3)\cdot k^\dagger$ \\
             & Conv ($\times$3)  & L1, 3, 5, Stride=2, $C_{\text{out}}=m^{*}\cdot k^\dagger$ \\
             & Conv ($\times$3)  & L2, 4, 6, Stride=1, $C_{\text{out}}=m^{*}\cdot k^\dagger$ \\
             & FC ($\times$2)    & 2000 LUTs per layer\\
    \midrule
    \multirow{4}{*}{FPGN-MLP} & FC ($\times4$) (CIFAR)      & 3000 LUTs per layer   \\
                              & FC ($\times4$) (KWS)        & 2048/1024/384/64 LUTs  \\
                              & FC ($\times3$) (JSC-OpenML) & 1000/1000/500 LUTs \\
                              & FC ($\times3$) (JSC-CERNBox)& 4000/3000/2000 LUTs \\                              
    \bottomrule
    \end{tabularx}
    \begin{tablenotes}[para,flushleft]
     \footnotesize 
     \item[$^*$] $m$ is set to 16 for Agg layer and 16, 32, 64 for different Conv layers.\\
     \item[$^\dagger$] $k$ is set to 1, 2, 4, 8 for -S, -M, -L, and -G, separately.
    \end{tablenotes}
\end{threeparttable}
\end{table}

\textbf{Inner-Level MILP Optimization.} Given a fixed set of $\{h_i\}$, the optimization of column-wise factors $\{w_i\}$ is linearized through a candidate-selection formulation. We define a binary decision variable $y_{i,j} \in \{0, 1\}$, where $y_{i,j}=1$ if the $j$-th configuration from a pre-profiled candidate set $\mathcal{P}_i$ is selected for layer $i$. The inner-level optimization is then formulated as an MILP:
\begin{equation}
    \begin{aligned}
        \min \quad & L_{\text{total}}=St_{\text{last}} + Pt_{\text{last}} \\
        \text{s.t.} \quad & \sum_{j \in \mathcal{P}_i} y_{i,j} = 1, \\
        & \sum_{i} \sum_{j \in \mathcal{P}_i} y_{i,j} \cdot R_{i,j}^{\text{type}} \le \text{Limit}_{\text{type}}, \quad \text{type} \in {\text{LUT, Reg}}
    \end{aligned}    
\end{equation}

By pre-calculating the resource costs and performance contributions for each candidate $j$ given a fixed $h_i$, the constraints and the objective function become linear combinations of $y_{i,j}$. This allows the compiler to utilize industrial solvers to find a conditionally optimal column-wise unroll factor within seconds. Combined with the heuristic search, this hybrid strategy ensures a high-quality configuration that balances spatial unrolling for lowest latency under the resource constraints, effectively mapping the LUT-native network onto the target FPGA fabric. 

Upon identifying the high-quality set of unroll factors, the compiler instantiates these parameters into the streaming architecture via a modular RTL-level template library. This automated process generates a fully synthesizable hardware implementation, effectively transforming the mathematical optimization results into a physical FPGA realization with minimal labor cost.

\begin{figure}
    \centering
    \includegraphics[width=0.9\linewidth]{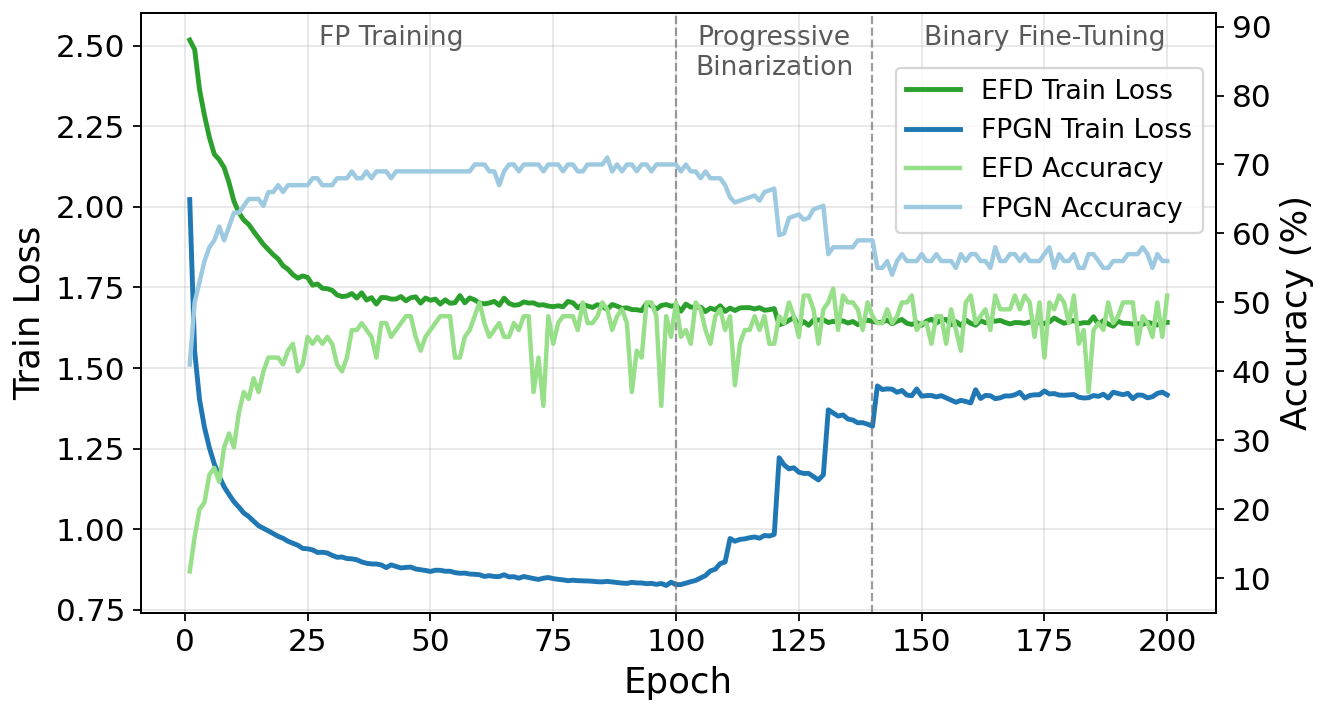}
    \caption{Training loss and accuracy trajectories. Compared to the discrete EFD rule, the proposed continuous formulation enables denser gradient updates (Epochs 1--100) and preserves forward-backward consistency during binarization (Epochs 100--200), leading to more effective convergence.}
    \label{fig_training_efficiency}
\end{figure}

\section{Evaluation}\label{sec:eva}
\subsection{Experimental Setups}

We evaluate the proposed FPGN framework across two architectural categories: (i) CNN-style FPGN-3/6 with various widths; and (ii) MLP-style configurations (FPGN-MLP). The detailed layer configurations are summarized in Table~\ref{tab_setup_compact}. These networks are trained from scratch on the CIFAR-10 \cite{CIFAR10}, SVHN \cite{netzer2011reading}, KWS \cite{warden2018speech}, and JSC (CERNBox version~\cite{jsccernbox} and OpenML version~\cite{jscopenml}) datasets using PyTorch on NVIDIA A30 GPUs. We compare FPGN against three representative paradigms: binary arithmetic centric (AMD official BNN framework FINN~\cite{umuroglu2017finn}), LUT-as-operator (LUTNet~\cite{wang2020lutnet}), and LUT-as-neuron (DWN~\cite{bacellar2024differentiable}, PolyLUT~\cite{andronic2023polylut}, NeuraLUT~\cite{andronic2024neuralut}, AmigoLUT~\cite{weng2025greater}, and NeuraLUT-Assemble~\cite{andronic2025neuralut}). For FPGA implementation, Post-P\&R metrics are obtained from AMD Vivado 2024.2, targeting the Versal Premium VP1902 platform (CNN-style networks) and Virtex UltraScale+ VU9P platform (MLP-style networks). The compiler is executed on an Intel i7-12700 CPU.

\subsection{Training Methodology Validation}\label{sec_eval_training}
We first validate our training methodology. This ensures that the subsequent architectural and physical results are grounded in an efficient and stable training process.

\subsubsection{Training Efficiency}

\begin{figure}
    \centering
    \includegraphics[width=0.9\linewidth]{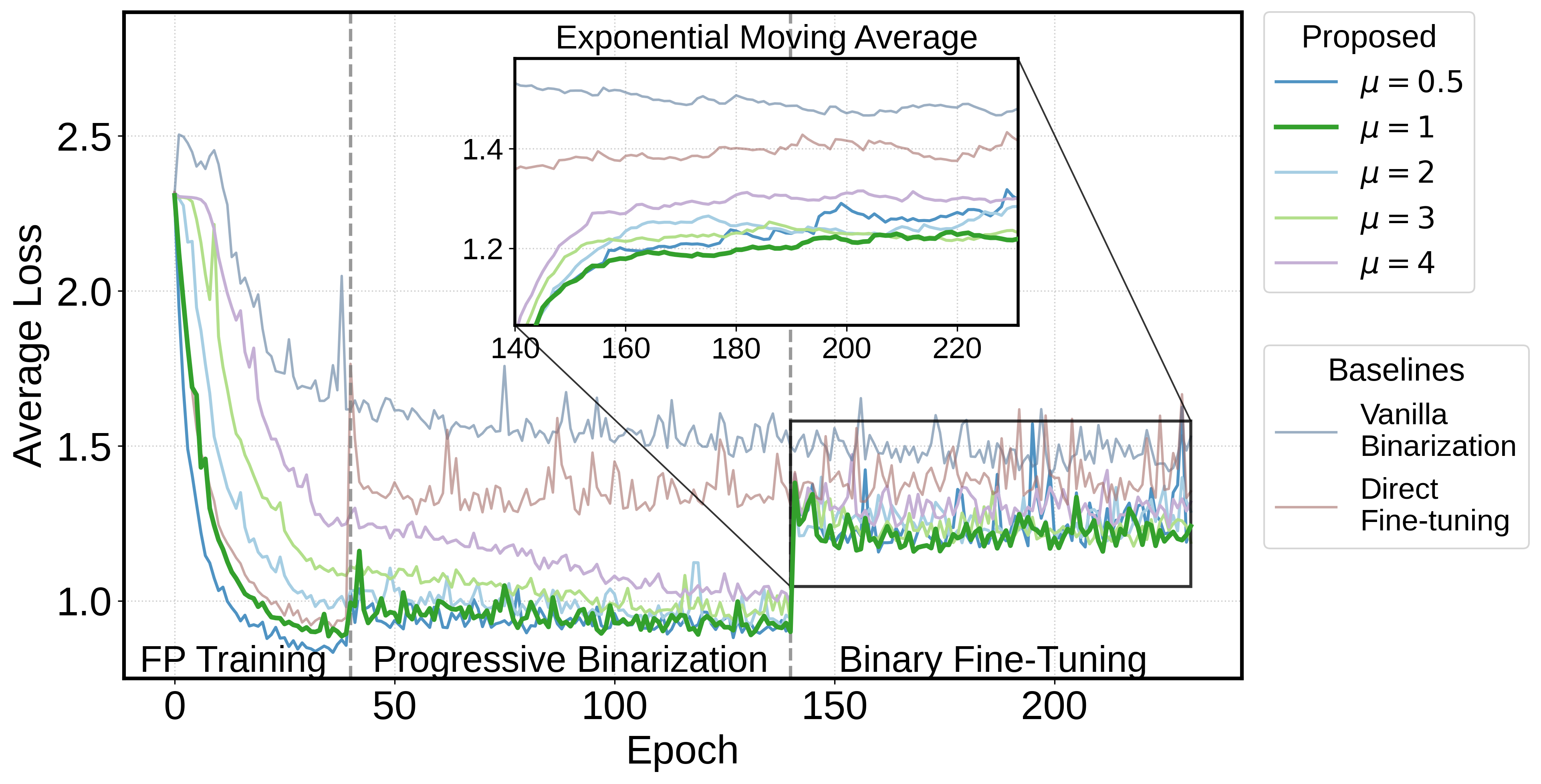}
    \caption{Training loss curves of FPGN under different initialization and staging configurations. Results show that the proposed three-stage strategy with $\mu=1$ (green) achieves the lowest final loss and best stability.}
    \label{fig_loss_curve}
\end{figure}

\begin{figure}[tbp]
    \centering
    \includegraphics[width=0.9\linewidth]{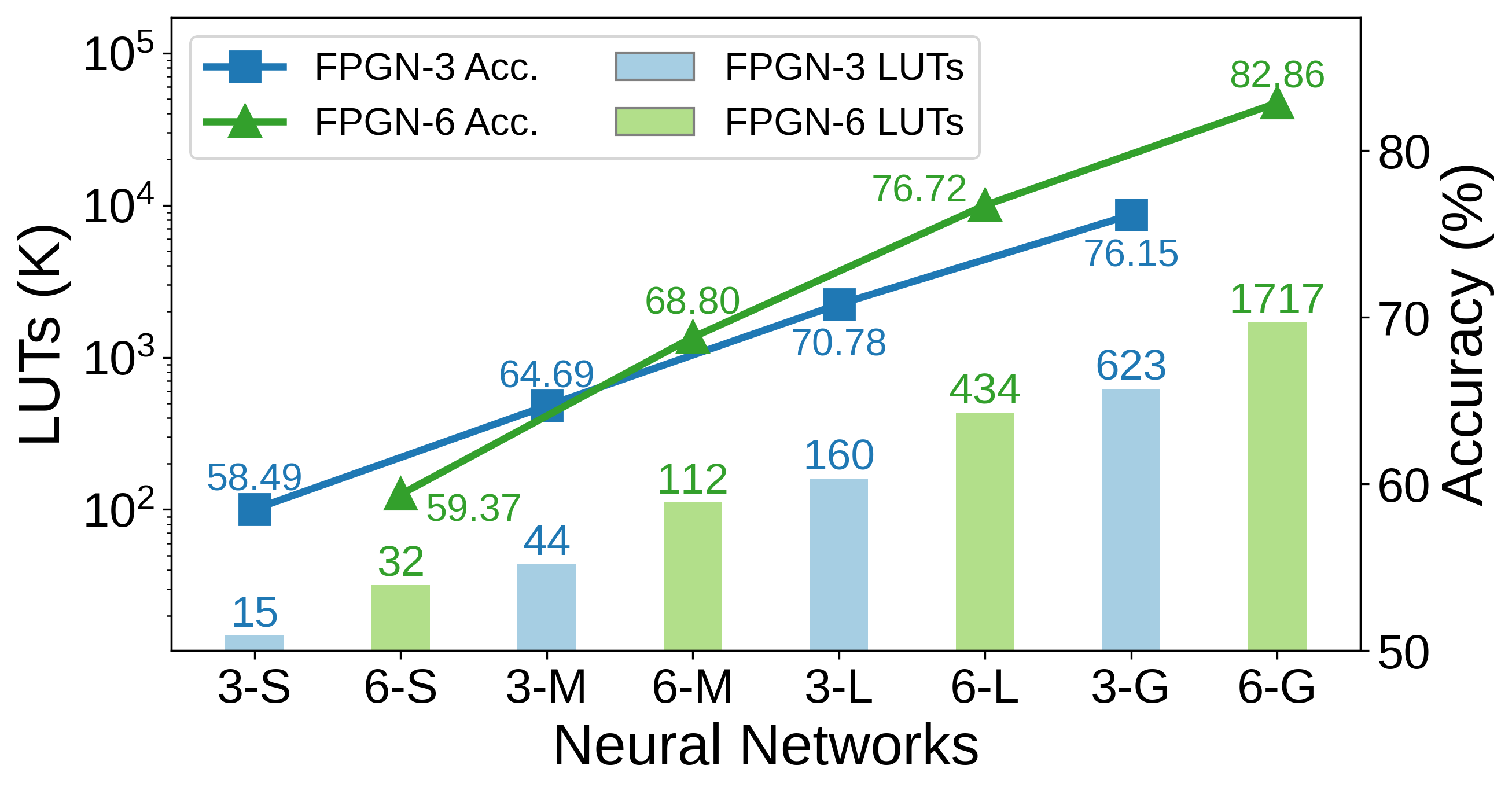}
    \caption{Accuracy and LUT usage of FPGN networks on CIFAR-10 dataset with different depths and scales.}
    \label{fig_compare_network}
\end{figure}

To evaluate the proposed continuous relaxation, we conduct an ablation study by comparing it with the discrete EFD method from DWN~\cite{bacellar2024differentiable}, the SOTA 6-LUT-based differentiable network, using FPGN-3-S on CIFAR-10. For a fair comparison, we deploy the EFD rule within FPGN's topology. As illustrated in Fig.~\ref{fig_training_efficiency}, the EFD baseline converges more slowly and exhibits larger fluctuations because its discrete perturbations restrict gradient updates to a single active entry out of the $2^k$ truth-table parameters per input sample. Conversely, during the relaxed optimization phase (Epochs 1--100), our continuous relaxation allows gradients to reach all $2^k$ LUT entries, yielding faster loss reduction and higher classification accuracy.

Additionally, EFD relies on an approximate gradient heuristic, which can introduce a mismatch between the discrete forward LUT logic and the backward optimization, which can reduce optimization effectiveness. In contrast, the proposed formulation recovers the exact finite-difference result at binary points. During progressive binarization and binary fine-tuning, FPGN maintains lower loss and higher accuracy than EFD, demonstrating more effective optimization under identical topological constraints.

\subsubsection{Ablation Study of Training Strategy}
We then evaluate the three-stage training strategy by comparing it against two baselines: vanilla binarization (trained entirely within the binary domain from scratch) and direct fine-tuning (immediately transitioning to the binary domain at Epoch 40, skipping the progressive stage). As illustrated in Fig.~\ref{fig_loss_curve}, Vanilla Binarization suffers from the highest loss and significant instability, as direct binary optimization from scratch leads to severe gradient fluctuations. While Direct Fine-tuning benefits from floating-point pre-training, the abrupt transition at Epoch 40 triggers a sharp loss spike that remains irrecoverable throughout the subsequent fine-tuning phase. In contrast, our methodology achieves significantly lower loss and superior stability, confirming that the progressive transition is essential for bridging the representation gap between continuous and binary states.

\begin{figure}
    \centering
    \includegraphics[width=0.9\linewidth]{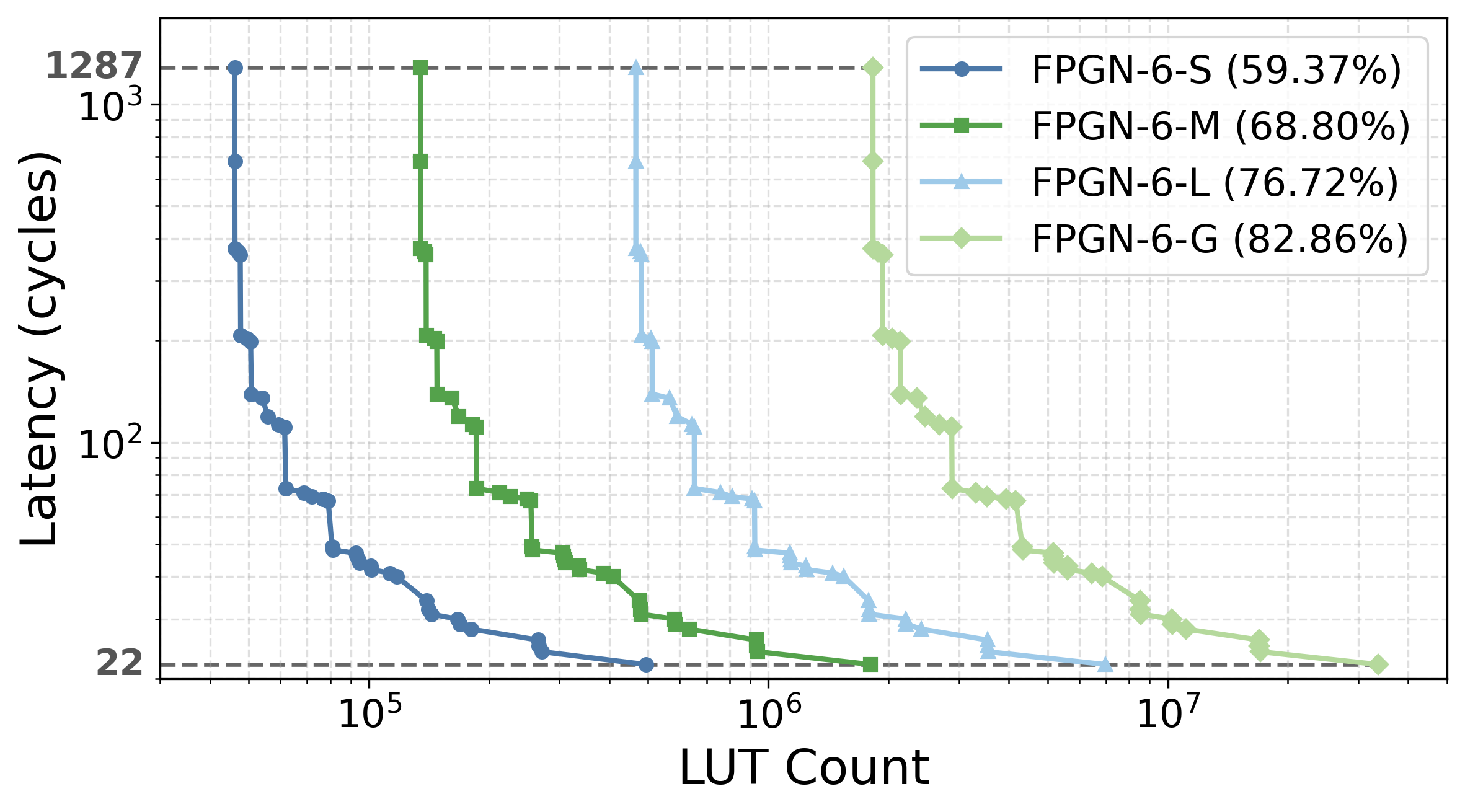}
    \caption{Compiler-estimated latency-LUT trade-off for FPGN-6-S/M/L/G under different unroll factors. S/M/L/G scale width to invest more LUTs for higher accuracy, while unroll factors trade LUTs for lower latency.}
    \label{fig_latency_lut}
\end{figure}

\begin{figure}
    \centering
    \includegraphics[width=0.9\linewidth]{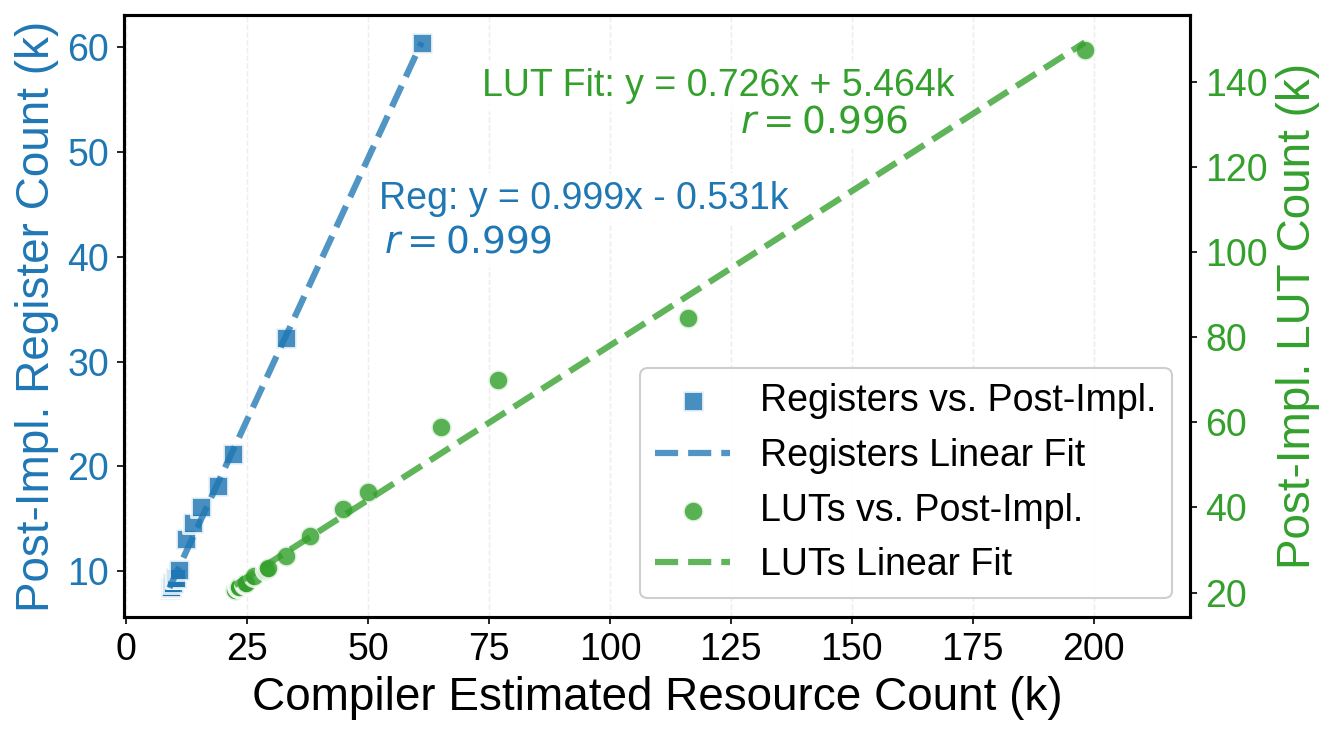}
    \caption{Correlation between compiler-estimated and Vivado post-implementation resources. The strict linear trend (Pearson $>0.99$) ensures the DSE algorithm correctly identifies optimal configurations.}
    \label{fig_model_correlation}
\end{figure}

\subsubsection{Ablation Study of Bimodal Initialization}
We further analyze the sensitivity of bimodal initialization by varying the mean value ($\mu$). Loss curves are shown in Fig.~\ref{fig_loss_curve}. For $\mu\ge2$, convergence during the FP training is hindered by vanishing gradients as weights reside within sigmoid saturation regions. In contrast, while $\mu=0.5$ facilitates rapid initial convergence, it causes weights to concentrate near the decision boundary (zero), leading to severe fluctuations during the progressive binarization and binary fine-tuning phases. As visualized in the exponential moving average (EMA) inset, this instability results in loss increase. Consequently, $\mu=1$ proves optimal, striking a balance between rapid initial convergence and robustness against binarization-induced perturbations.

\begin{table*}[tbp] 
    \caption{End-to-End Performance and Hardware Efficiency Comparison for CNN-Style Networks}
    \label{tab_master_eval}
    \centering
    \begin{threeparttable}
    \scriptsize
        \setlength{\tabcolsep}{3.5pt} 
        \begin{tabular}{@{} l c c r r r r r r r r r  c c c c @{}}
            \toprule
            \multicolumn{12}{l}{\textbf{Physical Metrics}} & \multicolumn{4}{l}{\textbf{Same Platform Norm. (rel. to FINN)}} \\
            \cmidrule(r){1-12} \cmidrule(l){13-16}
            \textbf{Method} & \textbf{Plat.} & \textbf{$F_\text{max}$} &\textbf{Acc.} & \textbf{LUT} & \textbf{FF} &\textbf{BRAM} & \textbf{Power$^\dagger$} & \textbf{Latency} & \textbf{FPS} & \textbf{FPS/LUT} & \textbf{kFPS/W} & \textbf{Latency} & \textbf{FPS} & \textbf{FPS/LUT} & \textbf{kFPS/W} \\
            \midrule
            \multicolumn{16}{l}{\textit{CIFAR-10 Dataset}} \\ 
            \midrule
            FINN    & KU115   & 228MHz & 80.1\% & 49.8k  & 67k & 116  & 3.3\,W  & 194\,$\mu$s      & 29.6k          & 0.59 & 8.97  & 1$\times$              & 1$\times$               & 1$\times$    & 1$\times$ \\
            LUTNet  & KU115   & 200MHz & 84.8\% & 106.8k & --  & --   & 5.4\,W  & --               & 10.2k          & 0.10  & 1.89  & --                     & 0.34$\times$            & 0.17$\times$ & 0.21$\times$ \\ 
            \midrule
            FINN    & VP1902  & 327MHz & 80.1\% & 49.6k  & 72k & 120  & 8.7\,W  & 135\,$\mu$s      & 42.5k          & 0.86 & 4.89  & 1$\times$              & 1$\times$               & 1$\times$    & 1$\times$ \\
            FPGN    & VP1902  & 205MHz & 82.9\% & 2.46M  & 924k& 0    & 89.8\,W & \textbf{658\,ns} & \textbf{3.21M} & 1.30  & 35.75 & \textbf{1/205$\times$} & \textbf{75.53$\times$}  & \textbf{1.51$\times$} & \textbf{7.31$\times$} \\
            \midrule
            \multicolumn{16}{l}{\textit{SVHN Dataset}} \\ 
            \midrule
            FINN    & KU115   & 228MHz & 94.9\% & 49.8k  & 67k & 116  & 3.3\,W   & 194\,$\mu$s & 29.6k   & 0.59 & 8.97  & 1$\times$     & 1$\times$     & 1$\times$    & 1$\times$ \\
            LUTNet  & KU115   & 200MHz & 96.4\% & 361.5k & --  & --   & --       & --          & 10.2k   & 0.03  & --    & --            & 0.34$\times$  & 0.05$\times$ & --  \\
            \midrule
            FINN    & VP1902  & 327MHz & 94.9\% & 49.6k  & 72k & 120  & 8.7\,W   & 135\,$\mu$s & 42.5k   & 0.86 & 4.89  & 1$\times$     & 1$\times$     & 1$\times$    & 1$\times$ \\
            FPGN    & VP1902  & 205MHz & 95.1\% & 2.46M  & 924k& 0    & 89.8\,W  & 658\,ns     & 3.21M   & 1.30  & 35.75 & 1/205$\times$ & 75.53$\times$ & 1.51$\times$ & 7.31$\times$ \\
            \bottomrule
        \end{tabular}
        \begin{tablenotes}[para,flushleft]
            \footnotesize
            \item --: Metric is not reported or cannot be directly derived from the available implementation/report.\\
            \item $\dagger$: Power follows the reported value for prior published results and Vivado-estimated on-chip power for re-implemented designs.\\
        \end{tablenotes}
    \end{threeparttable}
\end{table*}

\subsection{Topology Design Validation}
We validate the CNN-style topology that serves as the structural foundation for achieving massive spatial parallelism and ultra-low latency. The architecture allows for flexible scaling across both width and depth dimensions. We evaluate this scalability by comparing the shallower FPGN-3 and the deeper FPGN-6, across four scales (S, M, L, and G). As illustrated in Fig.~\ref{fig_compare_network}, both configurations achieve accuracy gains as the network scales in width, demonstrating the robustness of our topology under width-oriented expansion. While expanding layer width effectively enhances expressivity, our analysis reveals that depth scaling is fundamentally more resource-efficient than width-oriented expansion for LUT-native architectures. A primary observation is the comparison between FPGN-6-L and FPGN-3-G: the deeper FPGN-6-L outperforms the shallower but wider FPGN-3-G by 0.57\% in accuracy while utilizing 30\% fewer LUT resources. These results confirm the scalability of the proposed CNN-style topology. Combined with the structured connectivity analyzed in Sec.~\ref{sec:background}, the designed topology provides the necessary framework for high-frequency implementations.

This architecture also supports different unroll factors for latency optimization by the latency-driven compiler. Fig.~\ref{fig_latency_lut} illustrates a latency-resource trade-off under different unroll factors for FPGN-6. The four FPGN-6 variants scale model width to improve accuracy, at the cost of increased LUT usage. Since these FPGN-6 variants share the same depth and streaming schedule, they achieve the same latency under the same unroll factors, while wider variants require more LUTs. Therefore, smaller variants provide lower-resource deployment options with the expected accuracy trade-off, whereas larger variants reach higher accuracy under higher LUT budgets.

\subsection{Latency-Driven Compiler Validation}
\subsubsection{Compiler Fidelity}
Our compiler utilizes a LUT-centric analytical QoR model to navigate the vast design space. To validate its fidelity, we compare its estimates against Vivado post-P\&R results. For latency, since the FPGN architecture avoids dynamic scheduling, the predicted cycle counts are deterministic and precisely match the simulated hardware behavior. Regarding resource modeling, predicting post-P\&R utilization is inherently challenging due to aggressive logic optimizations (e.g., constant folding and retiming) by EDA tools. Thus, analytical models typically yield inaccurate absolute resource counts. However, for an automated DSE solver, the relative ordering is far more critical than absolute precision \cite{du2024fado,guo2021autobridge}. As demonstrated in Fig.~\ref{fig_model_correlation}, our resource estimation exhibits a near-perfect linear correlation with physical implementations, achieving Pearson correlation coefficients of 0.996 for LUTs and 0.999 for Registers. This exceptional fidelity stems from the deterministic nature of our LUT-native templates, which minimizes unpredictable mapping. By using the regression slope as a calibration factor, we systematically close the gap between analytical abstraction and physical synthesis, ensuring our MILP solver reliably converges to the true hardware optimum.

\subsubsection{DSE Solver Efficiency}
The proposed heuristic-MILP solver demonstrates high practical efficiency. When optimizing the complex FPGN-6-G network under resource constraints, a brute-force search over the vast unroll factor space would require days to complete due to the QoR model complexity. In contrast, our solver identifies a high-quality solution in only 16 seconds. This significant speedup confirms that our compiler provides an efficient and automated path from trained networks to optimized hardware implementations, enabling rapid design iteration for various architectures across diverse FPGA fabrics.

\subsection{End-to-End Evaluation}

Having validated each contribution independently, we now present a comprehensive end-to-end evaluation, comparing FPGN against representative SOTA works across three distinct paradigms. To ensure a fair comparison, We re-implemented the AMD Xilinx Official FINN on the same FPGA platform with LUTNet and FPGN as a normalization reference.

\subsubsection{Comparison with Arithmetic-Centric Paradigm} As motivated in Sec.~\ref{sec:intro}, FPGN prioritizes nanosecond-scale response over resource constraints. On CIFAR-10 datasets, FPGN-6-G achieves a deterministic latency of 658\,ns and a peak throughput of $3.21\,\text{M}$ FPS (Table~\ref{tab_master_eval}). Compared to the BNN baseline (FINN), FPGN delivers a 205$\times$ absolute latency reduction and 75.53$\times$ throughput increase.

This radical performance leap is enabled by a fundamental shift from sequential time-multiplexed execution to spatial investment of 2.46M LUTs. While FINN relies on time-multiplexed processing units and could theoretically improve performance by increasing parallelism, our normalized efficiency (FPS/LUT) reveals an inherent architectural gap. As shown in Table~\ref{tab_master_eval}, FPGN achieves 1.51$\times$ higher LUT efficiency than FINN. This implies that even if FINN utilized the same number of LUTs as FPGN, its throughput and latency would still fall significantly short of FPGN's. By treating LUTs as neurons, FPGN fully unleashes the fabric's Boolean expressive power. FFs in FPGN are primarily used for pipeline registers and inter-layer line buffers, while LUT utilization remains the dominant resource constraint. Since all trained parameters are embedded directly in LUT configurations, FPGN requires no BRAM for weight storage and DSP/AIE for computation. Together with its highly parallel streaming execution, this LUT-native datapath achieves \(7.31\times\) higher energy efficiency (kFPS/W) than the same-platform FINN baseline. This confirms that for ultra-low-latency regimes, the LUT-as-neuron paradigm is not only faster but also more power-efficient than the traditional arithmetic-centric paradigm.

\begin{table}
    \centering
    \caption{End-to-End Performance and Hardware Efficiency Comparison for MLP-Style Networks}
    \label{tab_mlp_comparison}
    \scriptsize
    \setlength{\tabcolsep}{3.5pt}
    \begin{tabular}{c l r r r r r} 
    \toprule
    \textbf{Dataset} & \textbf{Method} & \textbf{Acc.} & \textbf{LUT} & \textbf{REG} & \textbf{$F_\text{max}$} & \textbf{Latency} \\
    \midrule
    \multirow[b]{3}{*}{JSC} 
        & PolyLUT           & 75.1\% & 236541 & 12384 & 235\,MHz & 21.0\,ns \\ 
        & NeuraLUT          & 75.0\% & 92357  & 4885 & 368\,MHz & 14.0\,ns \\
        & AmigoLUT          & 74.4\% & 42742  & 4717 & 520\,MHz & 9.6\,ns \\
    \multirow[t]{3}{*}{CERNBox} 
        & NeuralLUT-        & \multirow{2}{*}{75.0\%} & \multirow{2}{*}{8539} & \multirow{2}{*}{1332} & \multirow{2}{*}{352\,MHz} & \multirow{2}{*}{5.7\,ns} \\ 
        & Assemble          &         &    &    &     &     \\ 
        & FPGN              & 74.9\%  & 12358 & 4839  & 669\,MHz & 6.0\,ns  \\
    \midrule
    \multirow[b]{2}{*}{JSC} 
        & DWN               & 76.3\% & 6302  & 4128& 695\,MHz & 14.4\,ns \\
        & NeuraLUT-         & \multirow{2}{*}{76.0\%} & \multirow{2}{*}{1780} & \multirow{2}{*}{540} & \multirow{2}{*}{941\,MHz} & \multirow{2}{*}{2.1\,ns} \\  
    \multirow[t]{2}{*}{OpenML} 
        & Assemble          &         &       &     &     & \\
        & FPGN              & 76.0\% & 3345    & 1703 & 730\,MHz  & 5.5\,ns \\
    \midrule
    \multirow{2}{*}{KWS}
        & DWN               & 71.5\% & 6169  & 1686 & 287\,MHz & 3.5\,ns  \\
        & FPGN              & 72.5\% & 3522  & 1607 & 407\,MHz & 2.5\,ns \\
    \midrule
    \multirow{2}{*}{CIFAR-10} 
        & DWN               & 57.4\% & 20837 & 3167 & 276\,MHz & 14.5\,ns \\
        & FPGN              & 58.1\% & 15336 & 2649 & 353\,MHz & 5.7\,ns \\
    \bottomrule
    \end{tabular}
\end{table}

\subsubsection{Comparison with LUT-as-Operator Paradigm} 
FPGN further demonstrates superior hardware efficiency compared to LUTNet, which also employs the 6-LUT fabric but represents the LUT-as-operator paradigm. While LUTNet replaces fixed XNOR gates with differentiable LUTs to increase expressive power, it retains the BNN datapath. As shown in Table~\ref{tab_master_eval}, although LUTNet utilizes LUT primitives, its hardware utility is severely constrained by this hybrid nature. On CIFAR-10, FPGN delivers a 222$\times$ throughput speedup with 9$\times$ higher LUT efficiency (FPS/LUT) over LUTNet. Unlike FINN and FPGN which share identical configurations across tasks, LUTNet incorporates heuristic pruning, leading to a varying model size on SVHN. On this dataset, FPGN's achieves a 30$\times$ higher LUT efficiency than LUTNet. This demonstrates that simply replacing binary operators with LUTs without rethinking the underlying topology is inherently inefficient, leaving LUTs' intrinsic expressivity and the fabric's LUT-native potential largely underutilized.

While LUTNet exhibits a marginally higher accuracy, this advantage relies on external orthogonal algorithmic heuristics, such as residual binarization~\cite{ghasemzadeh2018rebnet}, which FPGN intentionally excludes in comparison to isolate and validate the intrinsic raw efficiency of its LUT-native architectural paradigm. By applying 4-bit residual binarization, FPGN-6-L achieves a LUTNet-comparable accuracy of 84.42\% with a LUT budget similar to FPGN-6-G. Furthermore, similar to FINN, LUTNet suffers from the power overheads due to weight-related memory accesses. By contrast, FPGN delivers a substantial 34.8$\times$ energy efficiency improvement over LUTNet, proving that LUT-as-neuron paradigm is the more viable path for achieving nanosecond-scale neural acceleration on FPGA.

\subsubsection{Comparison with LUT-as-Neuron Paradigm} 
We evaluate FPGN-MLP variants against SOTA micro-scale LUT-as-neuron networks under identical hardware baselines. Table~\ref{tab_mlp_comparison} summarizes the results. Compared with DWN~\cite{bacellar2024differentiable}, the closest 6-LUT-based differentiable baseline, FPGN achieves comparable or higher accuracy across CIFAR-10, KWS, and JSC-OpenML dataset while using fewer LUTs. Its structured connectivity also enables lower latency. On CIFAR-10, FPGN and DWN consume $3.864\,\text{W}$ and $3.891\,\text{W}$, respectively, demonstrating comparable power consumption.

We further compare FPGN with function-parameterized MLP-style methods. As shown in Table~\ref{tab_mlp_comparison}, all works achieve competitive classification accuracy on two JSC versions. FPGN consumes fewer LUTs than PolyLUT, NeuraLUT, and AmigoLUT. This demonstrates that directly optimizing the global relaxation space of physical LUT entries expands the functional design space, thereby maximizing LUT expressivity. NeuraLUT-Assemble achieves higher resource efficiency on micro-scale networks. However, its reported end-to-end latency increases from $2.1\,\text{ns}$ on JSC-OpenML to $5.7\,\text{ns}$ on the more complex JSC-CERNBox workload. Since latency jointly reflects clock frequency and pipeline depth, this increase indicates reduced end-to-end latency scalability as the model size grows. In comparison, FPGN's structured connectivity is designed to preserve routing locality and latency scalability for larger deployments.

\subsubsection{Discussion}
Our primary comparisons focus on fully binary networks for ultra-low-latency inference. For broader context, a full-precision ResNet-20 achieves 91.25\% accuracy on CIFAR-10 with a measured batch-1 latency of \(9\,\text{ms}\) on an NVIDIA A30 GPU. Representative INT8/INT16 FPGA implementations retain approximately 90\% accuracy but operate at millisecond-scale latency~\cite{zhang2022wsq,qiu2016going}. Hybrid BNNs that preserve partial high-precision operations achieve approximately 86\%--88\% accuracy with representative FPGA latency in the hundreds-of-microseconds range~\cite{fayyazi2024neuroblend,zhang2021fracbnn}. Fully binary FPGA frameworks such as FINN operate at 80.1\% accuracy with $\sim200\mu\text{s}$ latency~\cite{umuroglu2017finn}. FPGN targets the extreme low-latency end of this design space, achieving 82.9\% accuracy with \(658\,\text{ns}\) latency. These representative results illustrate a general trend in which lower-precision paradigms trade accuracy for reduced inference latency. This study focuses on the fully binary operating regime rather than exhaustively exploring all intermediate precision and architecture configurations, which remains an important direction for future work.

Additionally, while following prior LUT-native networks to evaluate FPGN on CNN- or MLP-style architectures, FPGN’s primitives (LUT-Vector/Tree) map to matrix multiplication and reduction and can be applied to linear operations within Transformer architectures (e.g., QKV projections).  One future direction is to extend FPGN to more network architectures and more tasks.

\section{Conclusion}
In this work, we propose FPGN, a holistic co-design framework achieving nanosecond-scale neural acceleration via the novel LUT-as-neuron paradigm. By integrating hardware-aligned training, physically-aware topology, streaming hardware architecture, and latency-driven compilation, FPGN provides an end-to-end solution to translate algorithmic neurons to physical accelerators. Evaluations demonstrate a 205$\times$ latency reduction, alongside 30$\times$ higher LUT efficiency than SOTA LUT-as-operator methods. FPGN establishes a robust foundation for sub-microsecond neural acceleration, effectively unlocking the intrinsic logic speed of FPGA fabrics for the most demanding latency-critical applications.

{\small
\bibliographystyle{ieeetr}
\bibliography{reference}
}

\end{document}